\documentclass[
aps,pra,
twocolumn
]{revtex4}
\usepackage{graphicx}
\usepackage{float}
\usepackage{subfigure}

\usepackage{amsmath}
\usepackage{amssymb}
\usepackage{amsthm}
\usepackage{braket}
\usepackage{color}
\usepackage{microtype}
\usepackage{tikz}
\usepackage{makecell}
\usepackage{bm}

\usepackage{hyperref}
\hypersetup{colorlinks,linkcolor={blue},citecolor={red},urlcolor={magenta}}

\newcommand{\ba}{\begin{eqnarray}}
\newcommand{\ea}{\end{eqnarray}}

\theoremstyle{plain}

\theoremstyle{plain}

\begin{document}

\title{Mass distribution of ultralight boson in binary black hole systems}

\author{Hang Yang}
\email{hangyphy@hust.edu.cn}
\author{Daiqin Su}
\email{sudaiqin@gmail.com}

\affiliation{ MOE Key Laboratory of Fundamental Physical Quantities Measurement, Hubei Key Laboratory of Gravitation and Quantum Physics, PGMF, Institute for Quantum Science and Engineering, School of Physics, Huazhong University of Science and Technology, Wuhan 430074, China}

\date{\today}

\begin{abstract}

Ultralight bosons are compelling dark-matter candidates. Both scalar and vector bosons can be produced through black hole superradiance, forming a boson cloud surrounding a rotating black hole. Self-interaction of bosons, together with transition mixing in binary black hole systems, give rise to dynamical phenomena that could be potentially observable with future gravitational wave observations.
In this work, we investigate the dynamics of bosons in binary black hole systems. In particular, we focus on boson mass transfer in unequal-mass binary black hole systems with arbitrary spin-orientation of the companion. 
Our results show that the mass ratio between the companion and the primary black holes significantly affects cloud absorption through mass transfer. Moreover, when the companion's spin is not aligned with that of the primary, the efficiency of cloud depletion is further modified.

\end{abstract}

\maketitle

\section{Introduction}

The detection of gravitational waves by current ground-based~\cite{PhysRevLett.116.061102, Acernese_2015, PhysRevD.88.043007} and future space-borne~\cite{amaro2017laser, luo2016tianqin, hu2017taiji} gravitational wave detectors opens up a new avenue for exploring astrophysical processes that involve a strong gravitational field. One of the potential target sources for gravitational wave detectors is the ultralight boson, including axion and pseudoscalar axion-like particles. Axions are well-motivated solutions to the strong CP problem~\cite{PhysRevLett.38.1440, PhysRevLett.40.223, PhysRevLett.40.279}, and they are also potential candidates for dark matter~\cite{PRESKILL1983127, ABBOTT1983133, DINE1983137, PhysRevD.89.083536, PhysRevD.95.043541}. Cosmological observations on a wide range of scales provide promising avenues to search for them~\cite{Marsh:2015xka}. Axions and axion-like ultralight bosons can be produced through black hole superradiance~\cite{ZelDovich1971generation, ZelDovich1972Amplification, Misner1972Stability, Starobinski1973Amplification, Starobinski1974Amplification, PhysRevLett.119.041101, PhysRevD.96.024004, brito2020superradiance,Cromb_2020}, which leads to the formation of a boson cloud surrounding the black hole~\cite{PhysRevD.76.084001, PhysRevD.91.084011, Brito:2015oca, Baumann_2020}. The superradiance instability extracts mass and angular momentum from the black hole, thereby altering its evolutionary trajectory. Consequently, precision measurements of black hole masses and spins thus offer a powerful avenue to constrain the properties of ultralight bosons~\cite{PhysRevD.96.064050, Arvanitaki_2015, Arvanitaki_2010, PhysRevLett.126.151102, PhysRevD.103.083005, PhysRevLett.124.211101, PhysRevD.108.L041502, PhysRevD.108.124048, PhysRevLett.119.041101, PhysRevD.96.035019}. Moreover, the boson cloud can emit continuous gravitational waves due to its asymmetric spatial distribution and self-annihilation. With the detection capabilities of current and future gravitational wave observatories, gravitational waves offer a new window into astrophysical processes involving ultralight bosons~\cite{Li:2024rnk, Aggarwal:2020olq}. 

In a binary system, the time-dependent tidal field from the companion star can induce level mixing among different energy levels of the primary black hole, allowing bosons initially occupying a growing mode to transition into decaying modes, thereby depleting the boson cloud~\cite{PhysRevD.99.044001,PhysRevD.101.083019}. When the orbital separation becomes sufficiently small, the perturbation can also trigger ionization of the bosons~\cite{PhysRevD.105.115036}. Mixing of levels and ionization modify the orbital evolution of the binary system. In extreme cases, the companion may gain energy from the cloud at a rate that compensates gravitational wave emission, leading to the formation of a ``floating orbit"~\cite{PhysRevD.99.064018}. In contrast, the ionization of bosons could result in a rapid shrinkage of the orbit by extracting angular momentum. These processes substantially diversify gravitational wave signals~\cite{PhysRevLett.128.221102,Peng:2025zca}. When the companion is an S-star near the black hole, interactions with the high-density axion cloud may modify the effective fine-structure constant, thereby altering the stellar spectra~\cite{Bai:2025yxm}. If the companion is a pulsar, the resulting changes in orbital dynamics can be probed through pulsar timing observations~\cite{Ding_2021, Tong_2022}.

When the companion black hole moves sufficiently close to the primary, the gravitational potential of the companion can capture part of the bosons originally bound to the primary. As a result, the bosons may transfer from the primary to the companion, leading to a redistribution of the boson cloud. Beyond this classical gravitational capture, there is also the possibility of forming a bound gravitational ``molecule'', a configuration analogous to the hydrogen molecule~\cite{PhysRevD.101.124049, PhysRevD.103.024020}. Once the molecular state is established, the boson originally bound to the primary may occupy decaying modes around the companion if it is also a rotating black hole, thereby inducing further depletion of the boson cloud~\cite{guo2023masstransferbosoncloud}. 
For binaries in which the companion black hole penetrates the residual cloud surrounding the primary, this common envelope evolution can further influence the orbital separation and eccentricity~\cite{Guo:2025ckp,Guo:2024iye}, and the ionization process becomes important, leading to a characteristic turnover in the gravitational wave signals~\cite{Guo:2025pea}. 

In this work, we extensively study boson transfer and its depletion in a binary black hole system with several generalizations to Ref.~\cite{guo2023masstransferbosoncloud}. We begin by analyzing scalar bosons in binary black hole systems with unequal masses. In this circumstance, the bosons can be transferred to several orbitals of the companion black hole, instead of one in the case of equal mass, therefore giving rise to more decaying channels. 
Our results show that the boson cloud depletion is strongly dependent on the mass ratio. 
We then study the transfer and depletion of vector bosonic field in binary black hole systems. Since the vector bosonic field condensates in the ground state after superradiance, the companion black hole must move sufficiently close to capture the bosons from the primary. It has been shown that vector bosons hardly decay into the primary black hole due to Bohr mixing~\cite{Baumann_2020}. Here we show that vector bosons could be transferred to the companion black hole and completely decay into it, showing that mass transfer contributes dominantly to the cloud depletion. 
Finally, We investigate systems in which the spins of the two black holes are unparallel. Under the assumption that there is neither spin-spin coupling between two black holes nor spin-orbit coupling, we find that the dynamics of the scalar bosonic field is not affected by the relative spin orientation between the two black holes, while the vector bosonic field shows a reduced depletion fraction when the spin of the companion is tilted, highlighting the importance of spin geometry in determining the efficiency of vector-cloud depletion.

This paper is organized as follows. Section~\ref{Sec: Review} provides a brief review of scalar and vector boson clouds around a rotating black hole, including level mixing and ionization induced by tidal perturbations from the companion. In Sec.~\ref{Sec: Hypothesis}, we introduce the hypothesis of black hole ``molecular orbitals". Following this framework, Sec.~\ref{Sec: Scalar_cloud} applies the variational method to derive the eigenenergies and eigenstates of the scalar cloud, and examines their evolution and depletion.
Section~\ref{Sec: Vector_cloud} analyzes the vector-cloud eigenstates in binary black hole systems, and studies their corresponding evolution and depletion. In Sec.~\ref{Sec: Any_orientations}, we extend the analysis to arbitrary spin orientations of the companion and assess their impact on cloud depletion. Section~\ref{Sec: Gravitational_waves} discusses the resulting gravitational wave power. We summarize our findings in Sec~\ref{Sec: Conclusion}. 
Throughout this work, we use the natural unit with $G=\hbar=c=1$.

\section{Boson cloud in binary black hole system}
\label{Sec: Review}

In this section, we briefly summarize properties of the scalar and vector bosonic fields around a Kerr black hole. We first examine the superradiant conditions for both types of fields and then analyze the effects of perturbations induced by a companion black hole, which can trigger resonant transitions between bound states and ionization into the continuum states.

\subsection{Gravitational atom}

Suppose that a bosonic field with angular momentum $m$ and frequency $\omega$ interacts with a rotating black hole of mass $M$ and spin $J \equiv a M$. If the following condition is satisfied: 
\begin{eqnarray}
\frac{\omega}{m}<\Omega_{\rm{H}}=\frac{a}{2M r_+}, 
\end{eqnarray}
then the black hole (BH)-field system undergoes a superradiant instability, radiating bosons that extract energy and angular momentum from the black hole. Here, $m$ is the azimuthal angular momentum quantum number of the field, $\omega$ is the field's angular frequency, $\Omega_{\rm{H}}$ is the angular velocity of the black hole, and $r_+ = M+\sqrt{M^2-a^2}$ is the radius of the outer event horizon.

For both scalar and vector bosons, when the Compton wavelength of the boson exceeds the size of the black hole, bosons become non-relativistic and form hydrogen-like quasi-bound states. These states are denoted as \(\ket{\psi_{nlm}}\) for scalar bosons and \(\ket{\psi_{nljm}}\) for vector bosons, where \(n\) is the principal quantum number, \(l\) is the orbital angular momentum, and \(j\) is the total angular momentum. The combined system of the boson and the black hole is often referred to as a ``gravitational atom".
The radial profile of the wave function peaks at
\begin{eqnarray}
r_{c,n}\approx\Big(\frac{n^2}{\alpha^2}\Big)r_g=n^2 r_b,
\end{eqnarray}
where $r_g\equiv M$ is the gravitational radius of the black hole, and $r_b$ is the Bohr radius,  $\alpha=M\mu$ is a ``fine-structure like" constant. The eigenfrequencies of the bosons are given by~\cite{Baumann_2019}
\begin{eqnarray}\label{eq: eigenfrequencies}
\mathcal{E}_{nlm}^{\text{(scalar)}}&=& \mu\Big(1-\frac{\alpha^2}{2n^2}-\frac{\alpha^4}{8n^4}-\frac{f_{nl}}{n^3}\alpha^4+\cdots\Big)+i\Gamma_{nlm},\nonumber\\
\mathcal{E}_{nljm}^{\text{(vector)}}&=&\mu\Big(1-\frac{\alpha^2}{2 n^2}-\frac{\alpha^4}{8 n^4}+\frac{f_{n l j}}{n^3} \alpha^4+\cdots\Big)+i\Gamma_{nljm},\nonumber\\
\end{eqnarray}
where  
\begin{eqnarray}
f_{n l }&=&-\frac{6}{2l+1}+\frac{2}{n}, \\
f_{n l j}&=&-\frac{4(6 l j+3 l+3 j+2)}{(l+j)(l+j+1)(l+j+2)}+\frac{2}{n}. 
\end{eqnarray}
The key difference between the hydrogen atom and the gravitational atom lies in their boundary conditions. For the hydrogen atom, the boundary condition requires that the electron wave functions remain regular at the origin. In contrast, for a black hole, the presence of an event horizon imposes an inner boundary condition that could lead to radiation absorption by the black hole~\cite{PhysRevD.22.2323}. This implies that the boson may decay into the black hole, a process reflected by the imaginary part of the eigenfrequency. 
In the limit of $\alpha \ll 1$, the imaginary part $\Gamma$ of the scalar boson can be approximated analytically as~\cite{Baumann_2020}
\begin{eqnarray}
\Gamma_{nlm}=\frac{2r_+}{M}C_{nlm}(\alpha)(m\Omega_{\rm{H}}-\omega_{nlm})\alpha^{4l+5},
\end{eqnarray}
where $C_{nlm}$ is defined as
\begin{eqnarray}
C_{nlm}(\alpha)=\frac{2^{4l+1}(n+l)!}{n^{2l+4}(n-l-1)!}\Big[\frac{l!}{(2l)!(2l+1)!}\Big]^2\nonumber\\
\times \prod^l_{j=1}[j^2(1-\tilde{a}^2)+(\tilde{a}m-2\tilde{r}_+\alpha)^2],
\end{eqnarray}
with \(\tilde{a} \equiv a/M\) and \(\tilde{r}_+ \equiv r_+/M\). The energy levels with \(m \leq 0\) correspond to the decaying energy states. 
The instability rates for vector bosons are given by~\cite{Baumann_2019}
\begin{eqnarray}
\Gamma_{nljm}=2\tilde{r}_+ C_{nlj}g_{jm}(\tilde{a},\alpha,\omega)(m\Omega_{\rm{H}}-\omega_{nljm})\alpha^{2l+2j+5},
\end{eqnarray}
where the coefficients $C_{nlj}$ and $g_{jm}$ are given by
\begin{eqnarray}
C_{nlj}&=&\frac{2^{2l+2j+1}(n+l)!}{n^{2l+4}(n-l-1)!}\Big[\frac{l!}{(l+j)!(l+j+1)!}\Big]^2\nonumber\\
&&\times\Big[1+\frac{2(1+l-j)(1-l+j)}{l+j}\Big]^2,
\end{eqnarray}
and
\begin{eqnarray}
g_{jm}(\tilde{a},\alpha,\omega)= \prod_{k=1}^j\Big[k^2(1-\tilde{a}^2)+(\tilde{a} m-2r_+ \omega)^2\Big].
\end{eqnarray}
The superradiant condition for the vector fields applies to the total azimuthal angular momentum quantum number. Consequently, vector fields can exhibit superradiant growth even without orbital angular momentum, whereas the corresponding scalar modes are decaying. The dominant growing mode for ultralight vector fields is characterized by $n= 1$, $l = 0$, $j = 1$, and $m = 1$, which grows substantially faster than the leading scalar mode.

\subsection{Gravitational perturbations}

In a binary black hole system, when the separation between the two black holes is large, the influence of the companion black hole can be treated as a perturbation to the primary's Kerr metric~\cite{PhysRevD.99.044001}. By substituting the metric perturbation into the Klein-Gordon equation that governs the dynamics of the bosons, one finds that, to leading order, the metric deformation introduces perturbative terms in the effective potential for both scalar and vector bosons. For the scalar boson, the corresponding Schr\"{o}dinger equation can be written as 
\begin{eqnarray}
    i\frac{\partial }{\partial t}\psi(t,\mathbf{r})= \bigg(-\frac{1}{2\mu}\nabla^2-\frac{\alpha}{r}+\Delta V_{\rm scalar} \bigg)\psi(t,\mathbf{r}),
\end{eqnarray}
while for the vector boson, 
\begin{eqnarray}
    i\frac{\partial }{\partial t}\psi^m(t,\mathbf{r})=\bigg(-\frac{\delta^{mn}}{2\mu}\nabla^2-\frac{\alpha}{r}\delta^{mn}+\Delta V_{\rm vector}^{mn} \bigg)\psi^n(t,\mathbf{r}),
\end{eqnarray}
where $\psi^m$ represent the components of the vector wave function. 
These perturbations to the effective potential can be expressed as~\cite{PhysRevD.101.083019}
\begin{eqnarray}
    \Delta V_{\text{scalar}}&=&\frac{1}{2}\mu\delta g_{00},\nonumber\\
    \Delta V_{\text{vector}}^{mn}&=&\frac{1}{2}\mu\delta g_{00} \delta^{mn},
\end{eqnarray}
where $\delta g_{00}$ is one of the components of the metric perturbation. The multipole expansion of $\delta g_{00}$ is given by~\cite{Baumann_2019}
\begin{eqnarray}
  \delta g_{00}
=\sum_{l_*=2}\sum_{\left|m_*\right|\leq l_*}2M_*\varepsilon_{l_* m_*}(\Theta_*,\Phi_*)Y_{l_* m_*}(\theta,\phi)\nonumber\\
\times\Big[\frac{r^{l_*}}{R_*^{l_*+1}}u(R_*-r)+\frac{R_*^{l_*}}{r^{l_*+1}}u(r-R_*)\Big].
\end{eqnarray}
Here, $u(R_*-r)$ is the Heaviside step function, $\varepsilon_{l_* m_*}$ denotes the tidal moment induced by the companion along the geodesic of the BH-cloud, 
and $(t, x)$ refers to the center-of mass of the unperturbed BH-cloud in comoving Fermi coordinates.
We denote the position of the companion relative to the BH-cloud frame as $\mathbf{R}_*(t)\equiv\{R_*(t), \Theta_*(t), \Phi_*(t)\}$.

The angular perturbations induced by the gravitational disturbance are proportional to the spherical harmonics $Y_{l_* m_*}(\theta,\phi)$, leading to an overlap between two orbits $\psi_a$ and $\psi_b$ of the form
\begin{eqnarray}
    \bra{\psi_a}V_*\ket{\psi_b} \sim \int Y_{l_* m_*}Y^*_{l_a m_a}Y_{l_b m_b} \sin \theta {\rm d} \theta {\rm d} \phi. 
\end{eqnarray}
According to the time-dependent perturbation theory, a non-vanishing overlap $\bra{\psi_a}V_*\ket{\psi_b}$ induces transitions between the corresponding two states. Suppose that $\ket{\psi_b}$ is a bound state. If $\ket{\psi_a}$ is also a bound state, this transition corresponds to a resonant transition~\cite{PhysRevD.99.044001}. While if $\ket{\psi_a}$ belongs to the continuum, implying that the boson is unbound, the process becomes analogous to the photoelectric effect and is commonly referred to as ionization~\cite{PhysRevD.105.115036}.

\subsubsection{Resonant mixing}

For the scalar field, the dominant growing mode is $\ket{211}$. During the inspiral of the binary, the companion black hole induces hyperfine mixing between the growing mode $\ket{211}$ and the lower-energy decaying mode $\ket{21{-}1}$, as well as Bohr resonance between $\ket{211}$ and higher-energy decaying modes with $n \ge 3$~\cite{PhysRevD.99.044001,Baumann_2020}. The transition probability from the growing mode to the decaying mode is given by
\begin{eqnarray}
    |C_d(t)|^2=\bigg[1-\Big(\frac{\epsilon\mp\Omega}{\Delta_R}\Big)^2\bigg]\sin^2\bigg[\int_{t_0}^t {\rm d} t' \Delta_R(t') \bigg].
\end{eqnarray}
Here, $\epsilon$ denotes the energy gap between the growing and decaying modes, $\Omega$ is the angular frequency of the binary black holes, and $\eta$ represents the strength of the gravitational perturbation, and
\begin{eqnarray}\label{eq:detuning}
    \Delta_R=\sqrt{\eta^2+(\epsilon \mp \Omega)}\,.
\end{eqnarray}
The minus sign in Eq.~\eqref{eq:detuning} corresponds to counter-rotating orbits, while the plus sign corresponds to co-rotating orbits. 
The resonance occurs when $\epsilon \mp \Omega = 0$, from which one can determine the resonant orbital separations for hyperfine mixing and Bohr mixing. 
Since the energy gap associated with hyperfine mixing is narrower than that of Bohr mixing, hyperfine resonance occurs at much larger orbital separations.
In the co-rotating binary case, Bohr mixing does not occur, and hyperfine mixing dominates the cloud depletion. While in a counter-rotating binary, hyperfine resonance is absent and the Bohr mixing dominates the cloud depletion.

\begin{figure}[htbp]
    \centering
    \includegraphics[width=1\columnwidth]{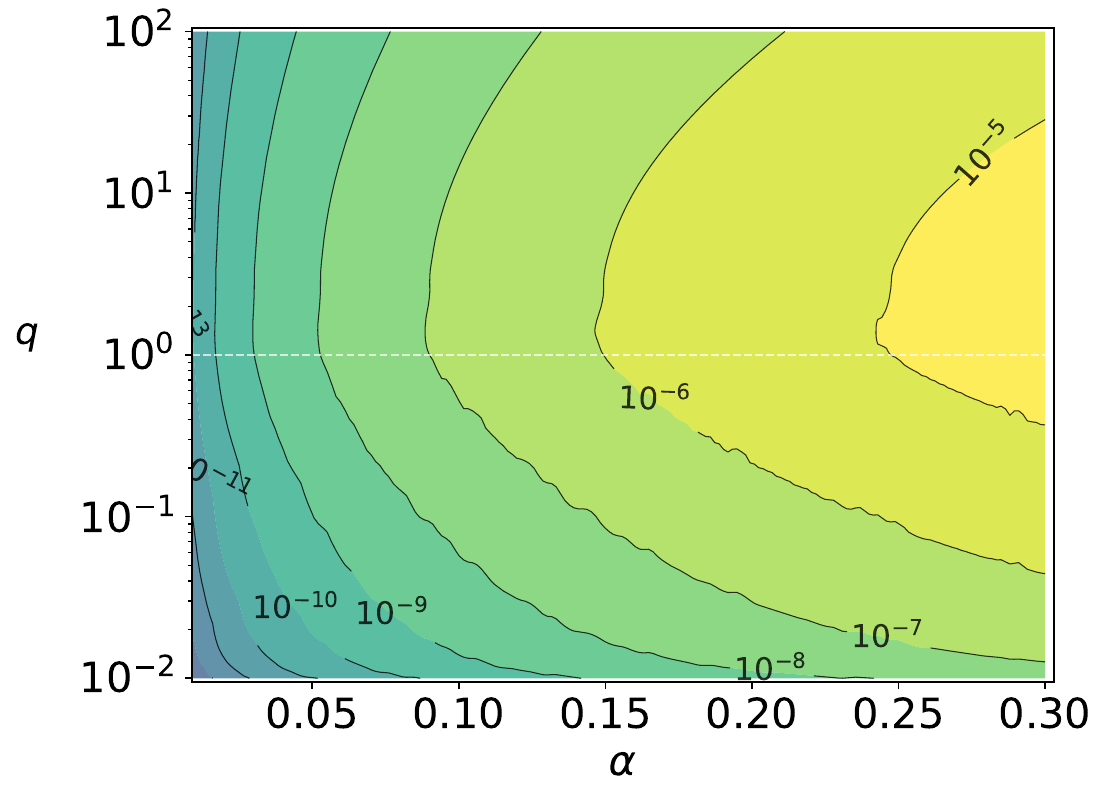}
    \caption{Fraction of the initial vector cloud mass that depletes into the primary black hole via Bohr mixing. It is assumed that the inspiral of the binary black hole begins at an orbital separation of $R_*=1000 r_b$ and evolves down to $R_*=3r_b$, where the ionization process starts to dominate. } 
    \label{fig:depletion into primary}
\end{figure}

For the vector field, the dominant growing mode is $\ket{1011}$. The hyperfine resonance associated with this mode is forbidden, leaving only the Bohr resonance allowed~\cite{PhysRevD.101.083019}, which can be excited in a counter-rotating binary black hole system. The nearest decaying mode is $\ket{321{-}1}$; however, the resulting depletion of the boson cloud through this channel is extremely weak and can be neglected~\cite{Baumann_2020}. 
Figure~\ref{fig:depletion into primary} shows the fraction of the initial mass of the boson cloud that is depleted into the primary black hole via Bohr mixing as a function of the mass ratio $q$ and the fine structure constant $\alpha$. Across most of the parameter space, this fraction remains below $10^{-5}$, indicating that level mixing is an inefficient depletion mechanism for the vector bosonic field. It should be noted that the self-interaction of the vector field is not included in this estimate.

\subsubsection{Ionization}

The gravitational perturbation from the companion acts like a quasi-periodic driving force with instantaneous frequency $\Omega(t) = |\dot{\varphi}_*(t)|$, where $\varphi_*(t)$ is the companion's polar angular relative to the primary black hole in the equatorial plane. When the driving frequency is sufficiently high, the perturbation can induce resonance transitions between bound states and continuum states, a process known as ionization. Suppose that the bound state is denoted as $\ket{n_b, \ell_b, m_b}$ and has a negative energy $\epsilon_b$, and the continuum state is denoted as $\ket{k_*, \ell, m}$, with $k_*$ the momentum of the boson. In the non-relativistic limit, $k_* \ll \mu$, the energy of the boson is approximated as $k_*^2/2\mu$. The ionization occurs when the following condition is satisfied~\cite{PhysRevD.105.115036}, 
\begin{equation}
\frac{1}{2\mu}k_*^2(t) = g \dot\varphi_* + \epsilon_b,
\end{equation}
where $g = m - m_b \in \mathbb{Z}$, known as the overtone, is the difference of azimuthal angular momenta between the continuum and bound states. In general, ionization occurs later than the resonance transitions between bound states, e.g., the hyperfine and Bohr mixings. Due to the presence of multiple overtones in the gravitational perturbation, resonances can occur at multiple distinct radii, which are given by~\cite{PhysRevD.105.115036}
\begin{eqnarray}
R_*^{(g)}=[4g^2(1+q)n_b^4]^{1/3}\alpha^{-2}r_g.
\end{eqnarray}
When the companion black hole enters the boson cloud, a gravitational molecule is formed, causing the bosons to co-move with the black holes and thereby modifying the ionization process~\cite{Guo:2025pea}. However, since we focus only on the characteristic scale at which ionization occurs, we restrict our analysis to the simplest case.

\section{Formation of molecular orbitals}
\label{Sec: Hypothesis}

We study how the presence of a companion black hole leads to the redistribution of bosonic field in the binary black hole system via formation of molecular orbits, and the subsequent depletion of the boson cloud into the companion black hole. We relax the assumption of equal mass that was studied in Ref.~\cite{guo2023masstransferbosoncloud} and consider an arbitrary mass ratio, which results in formation of distinct molecular orbits and thus gives rise to different depletion channels. In this section, we assume that the companion moves in the equatorial plane of the primary black hole, and the spins of these two black holes align and are perpendicular to the equatorial plane. We consider the case for an arbitrary spin orientation of the companion black hole in Sec.~\ref{Sec: Any_orientations}.

\subsection{Born-Oppenheimer approximation}

When the boson's relaxation time~\cite{guo2023masstransferbosoncloud} $\tau_r\sim R/\alpha$
is much shorter than the orbital period of the binary black holes, the effect of rotation can be neglected. The Hamiltonian of the boson is approximated as the sum of its kinetic energy and the potential energy in the gravitational field of the black holes,
\begin{eqnarray}\label{eq:Hamiltonian}
H=-\frac{\nabla^2}{2\mu}-\frac{\alpha_1}{r_1}-\frac{\alpha_2}{r_2},
\end{eqnarray}
where $r_1$ and $r_2$ denote the distances from the boson to the primary and companion black holes, respectively; while $\alpha_1$ and $\alpha_2$ are the corresponding fine-structure-like constants. In general, the orbital period of the binary system is much shorter than the timescale of coalescence. Based on the above two approximations, it is reasonable to introduce instantaneous molecular orbits, the wavefunctions of which vary slowly as the orbital separation shrinks due to the emission of gravitational waves.

\subsection{Formation rules of molecular orbits}

Without loss of generality, we assumed that initially the boson cloud exists only around the primary black hole and not around the companion. The subsequent analysis can be straightforwardly generalized to the case in which both black holes host their own boson clouds. 
When two black holes are widely separated, the boson cloud remains localized around the primary black hole. As the binary gradually approaches, a portion of the boson cloud enters the gravitational potential of the companion, allowing the companion to capture some of the bosons. This redistribution of the bosons is analogous to the formation of molecular orbitals in the hydrogen molecular ion. The exact wavefunctions for these molecular orbitals can be obtained by solving the time-dependent Schr\"odinger equation governed by the Hamiltonian in Eq.~\eqref{eq:Hamiltonian}. 

The variational method provides a simple semi-analytic approach for obtaining approximate solutions to molecular orbitals and has been widely applied in physical chemistry. In this method, the molecular orbitals are assumed to be linear superpositions of the atomic orbitals associated with the two nuclei, and the coefficients are determined by seeking stationary solutions of the expectation value of the Hamiltonian. The construction of molecular orbitals follows several key principles~\cite{lee2020concise}:
\begin{itemize}
    \item  The associated atomic orbitals must have similar energies to enable strong bonding interactions.
    \item  The orbitals of the primary black hole must overlap effectively with those of the companion, which requires the two black holes to be sufficiently close. 
    \item  The molecular orbitals must reflect the symmetry of the system's Hamiltonian, implying that the contributing atomic orbitals must either remain unchanged under rotation about the internuclear axis or transform identically under such rotations.
\end{itemize}
The binary black holes continuously emit gravitational waves so their orbit gradually shrinks, resulting in time-varying orbital separation and wave-function overlaps. During the timescale within which the orbital separation does not have significant changes, instantaneous molecular orbits can be introduced to study the transfer of bosons between the two black holes.

\section{Scalar cloud}
\label{Sec: Scalar_cloud}

In this section, we study the dynamics of scalar bosons in a binary black hole system with unequal masses. Since the dominant growing mode of the scalar boson generated by superradiance is the $\ket{211}$ mode, we assume that the primary black hole is initially surrounded by a boson cloud with all bosons occupying the state $\ket{211}$. We focus on the dynamical evolution of an already populated scalar cloud and do not model the superradiant growth phase explicitly. The influence of the companion black hole on the growing energy levels mentioned in~\cite{Fan:2023jjj,Li:2025qyu} is therefore neglected. We also ignore the effects of boson self-interactions \cite{Takahashi:2024fyq,PhysRevD.103.095019} and multi-field dynamics \cite{Zhu:2025enp}, assuming that they provide only subleading corrections in the regime of interest. To simplify the analysis, it is further assumed that initially the companion black hole is empty of bosons. In the subsequent evolution of the binary black hole, bosons could transfer from the primary black hole to the companion black hole. According to the Hamiltonian in Eq.~\eqref{eq:Hamiltonian}, the system possesses rotational symmetry about the $x$-axis. Based on the three formation rules outlined above, molecular orbitals can be classified according to their symmetry: $\sigma$ orbitals exhibit rotational symmetry about the x-axis, while $\pi$ molecular orbitals are symmetric in the x-y plane and acquire a minus sign under a  $180^\circ$ reflection about the $x$-axis. The $\pi$ orbitals can be constructed from $p$ orbitals,  
as well as higher atomic orbitals with larger principal and orbital quantum numbers, such as $\ket{\psi_{32-2}}$. However, in this work, we restrict ourselves to the case of $l=1$, for two reasons: (i) the wave functions with $l=1$ have a larger overlap with those of the primary black hole, and (ii) the decay rate of atomic orbitals with magnetic quantum number $m=-1$ is significantly higher than that of orbitals with $m<-1$. As a result, $\sigma$ orbitals can be formed only from $s$ and $p$ states, whereas $\pi$ orbitals can arise from $p$ states.

\begin{figure}[htbp]
    \centering
    \includegraphics[width=1\linewidth]{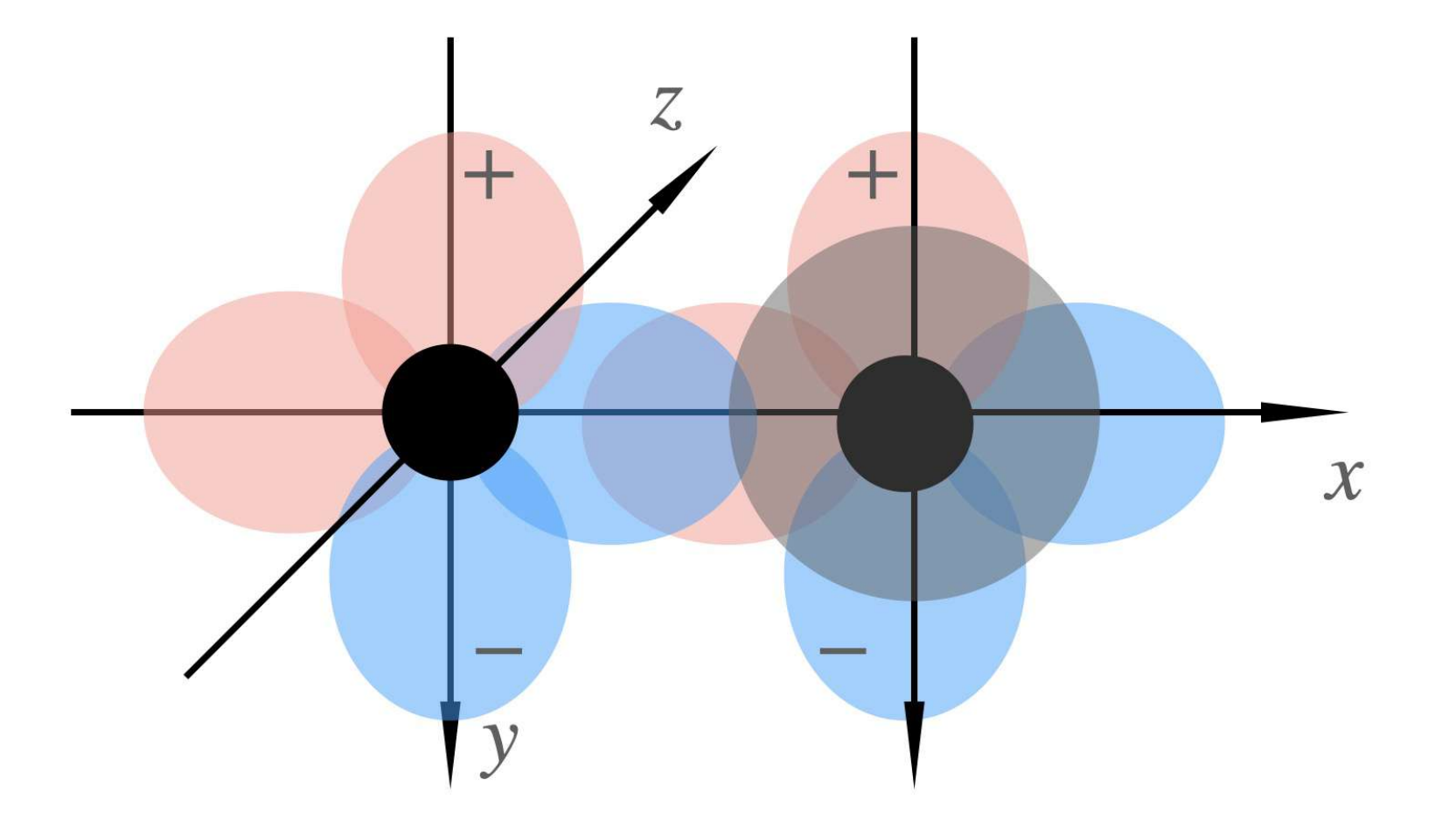}
    \caption{Illustration of a binary black hole system: The central black hole is surrounded by a scalar boson cloud occupying the states $\ket{\psi^1_{2p_x}}$ and $\ket{\psi^1_{2p_y}}$. The companion black hole depletes the cloud through its $s$- and $p$-states. Pink regions indicate positive components of the wavefunction,  blue regions indicate negative components, and the gray sphere represents the $s$-state. As the companion black hole approaches, the $\ket{\psi^1_{2p_x}}$ state begins to overlap with both the companion's $s$- and $p_x$-state. In contrast, $\ket{\psi^1_{2p_y}}$ overlaps effectively only with the companion's $p_y$-state; it positive and negative components calcel out when interacting with companion's $s$-state, resulting in no net depletion from that channel.}
    \label{fig:binary with 2px2py}
\end{figure}

When constructing ansatz wave functions in the variational method, it is convenient to consider wave functions with definite spatial symmetries, e.g., $p_x$ and $p_y$ states. 
\begin{eqnarray}\label{eq:npxnpy}
 \ket{\psi_{np_x}}&=& \frac{1}{\sqrt{2}}(\ket{\psi_{n1-1}}-\ket{\psi_{n11}}),\nonumber\\
 \ket{\psi_{np_y}}&=&\frac{i}{\sqrt{2}}(\ket{\psi_{n11}}+\ket{\psi_{n1-1}}),
\end{eqnarray}
where $n=1, 2, 3\cdots$. In the following discussion, we use superscripts ``1" and ``2" to denote states associated with the primary and companion black holes, respectively.
It turns out that the state \(\ket{\psi^1_{2p_x}}\) overlaps with both \(\ket{\psi^2_{ns}}\) and \(\ket{\psi^2_{np_x}}\), whereas \(\ket{\psi^1_{2p_y}}\) has zero overlap with \(\ket{\psi^2_{ns}}\). Therefore, we can construct two different types of ansatz wave functions. The $\sigma$ orbitals are formed from both $s$ and $p$ states, while the $\pi$ orbitals are constructed exclusively from $p$ states. 
\begin{eqnarray}\label{eq: scalar molecular orbits} 
\ket{\sigma_s}&=&c_{a}\ket{\psi^1_{2p_x}}+\sum_{n=1}^N( c_{ns}\ket{\psi^2_{ns}}+c_{np_x}\ket{\psi^2_{np_x}}),\nonumber\\
\ket{\pi_s}&=&c_{b}\ket{\psi^1_{2p_y}}+\sum_{n=1}^N c_{np_y}\ket{\psi^2_{np_y}}.
\end{eqnarray}

Here, the states denoted by $\sigma$ exhibit rotational symmetry about the $x$-axis, whereas the $\pi$ states are symmetry in the $x$-$y$ plane and acquire a minus sign under a $180^\circ$ rotation about the $x$-axis. The subscript $s$ is used to indicate that these molecular orbitals belong to the scalar cloud. 

To simplify notation, we use $\ket{\Phi}$ to denote either $\ket{\sigma_s}$ or $\ket{\pi_s}$, and $\{ \ket{\Psi_i} \}$ to represent a set of  isolated atomic orbits for both black holes. For $\ket{\Phi} = \ket{\sigma_s}$, the basis states $\ket{\Psi_i}$ include \{$\ket{\psi^1_{2p_x}}$, $\ket{\psi^2_{ns}}$, $\ket{\psi^2_{np_x}}$\}; whereas for $\ket{\Phi} = \ket{\pi_s}$, the basis states $\ket{\Psi_i}$ are \{$\ket{\psi^1_{2p_y}}$, $\ket{\psi^2_{np_y}}$\}.
Our goal is to find an approximation to the energy eigenvalues and eigenstates of the Hamiltonian in~\eqref{eq:Hamiltonian}. Suppose $\ket{\Phi}$ is the eigenstate of the Hamiltonian with an eigenvalue $E$, then we have 
\begin{eqnarray}\label{eq:energy_eigen_value}
\bra{\Psi_i}H\ket{\Phi}=E\braket{\Psi_i | \Phi}.
\end{eqnarray}
Taking into account the principal quantum number of the companion from $n=1$ to $n=N$, the substitution of the ansatz in Eq.~\eqref{eq: scalar molecular orbits} into Eq.~\eqref{eq:energy_eigen_value} yields a total of $2N$ equations for the $\sigma$ orbitals and $N$ equations for the $\pi$ orbitals. For $\sigma$ orbitals, 
\begin{eqnarray}\label{eq: solve_coefficients}
    (H_{i1}-E S_{i1})c_{a}+\sum^N_{j=1}(H_{ij}-E S_{ij})c_{ns}\nonumber\\
    +\sum_{k=2}^N(H_{i(N+k-1)}-E S_{i(N+k-1)})c_{np_x}=0,
\end{eqnarray}
where the index $i$ runs from $1$ to $2N$. The quantity $S_{ij}=\braket{\Psi_i \mid \Psi_j}$ denotes the overlap integral between two wave functions, and $H_{ij}=\bra{\Psi_i} H \ket{\Psi_j}$ represents the corresponding matrix elements of the Hamiltonian $H$. Since the wave functions $\ket{\Psi_i}$ are real, it follows that $S_{ij}=S_{ji}$ and $H_{ij}=H_{ji}$. A similar set of equations can be obtained by substituting the $\pi$ orbitals into Eq.~\eqref{eq:energy_eigen_value}. 

The energy eigenvalue $E$ and its corresponding eigen-state, which is characterized by coefficients $\{ c_i \}$, are unknown and must be solved from Eq.~\eqref{eq: solve_coefficients}. Non-trivial solutions for the coefficients exist only when the determinant of the coefficient matrix vanishes, 
\begin{eqnarray}\label{eq:determinant}
    \begin{vmatrix}
    H_{11}-E  &\cdots & H_{1(2N)}-E S_{1(2N)} \\
    H_{21}-E S_{21} &\cdots & H_{2(2N)}-E S_{2(2N)}\\
    \vdots & \ddots & \vdots\\ 
   H_{(2N)1}-E S_{(2N)1} &\cdots & H_{(2N)(2N)}-E
    \end{vmatrix}=0.
\end{eqnarray}

It is difficult to write explicit solutions for the energy eigenvalue $E$ from Eq.~\eqref{eq:determinant}. Instead, we numerically solve Eq.~\eqref{eq:determinant} for $E$ and then find its corresponding set of coefficients $\big( f_{a0}, f_{an}, f_{a(N+n-1)} \big)$ by substituting the value of $E$ into Eq.~\eqref{eq: solve_coefficients}. There are multiple solutions for $E$ and their corresponding orbitals are abbreviated as $\sigma^a$,
\begin{eqnarray}
\ket{\sigma^a_s}&=&f_{a0}\ket{\psi^1_{2p_x}}+\sum_{n=1}^{N}f_{an}\ket{\psi^2_{n s}} \nonumber\\
&&
+\sum_{n=2}^N f_{a(N+n-1)}\ket{\psi^2_{n p_x}}, 
\end{eqnarray}
where $a=1,2,3,\cdots$, which labels different 
$\sigma$ molecular orbitals. 
The limit of the summation index $1$ to $N$ indicates that the associated principal quantum number is considered from $1$ to $N$. 
Similarly, one can derive the energy eigenvalues and their corresponding coefficients $(g_{\mu 1}, g_{\mu n})$ when $\ket{\Phi} = \ket{\pi_s}$. Their corresponding orbitals are abbreviated as $\pi_\mu$, 
\begin{eqnarray}
\ket{\pi^\mu_s}=g_{\mu 1}\ket{\psi^1_{2p_y}}+\sum_{n=2}^{N}g_{\mu n}\ket{\psi^2_{n p_y}},
\end{eqnarray}
where $\mu=1, 2, 3, \cdots$, which denotes different $\pi$ molecular orbitals.

\begin{figure}[htbp]
    \centering
    \includegraphics[width=1\columnwidth]{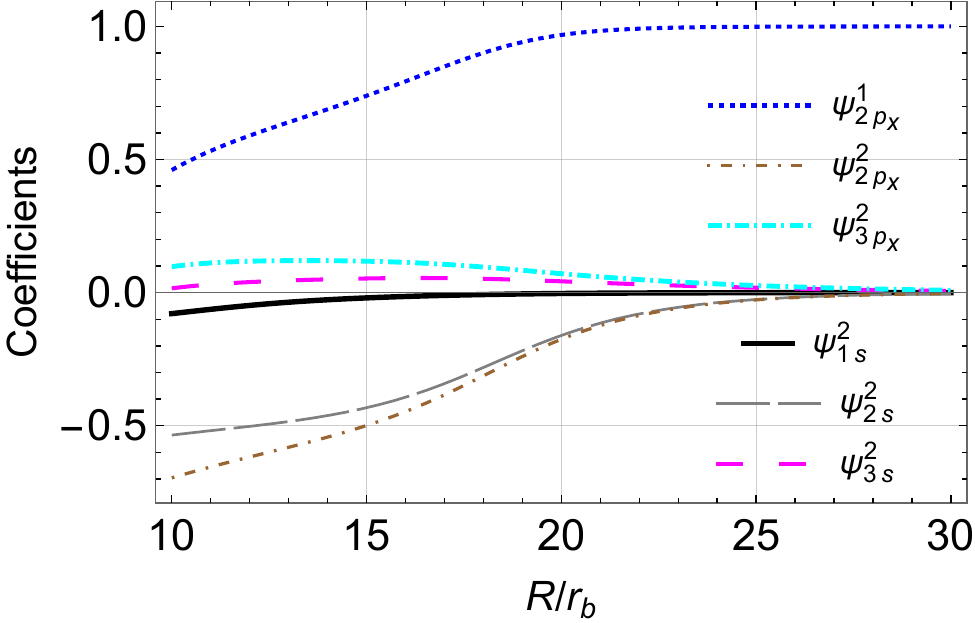}
    \caption{Coefficients of the molecular orbitals corresponding to $E_1$, where we choose $q=0.9$ and fine-structure-like constant $\alpha=0.1$.}
    \label{fig:coefficients}
\end{figure}

Since $S_{ij}$ and $H_{ij}$ are functions of the orbital separation of the two black holes, both energy eigenvalues and their corresponding coefficients depend on the orbital separation. Initially, when two binary black holes are widely separated, the wave function overlaps $S_{ij}$ and the off-diagonal elements of the Hamiltonian vanish. In this limit, the energy eigenvalues are reduced to $H_{ii}$, which corresponds to the eigenvalue of the isolated states of an individual black hole. As the two black holes gradually approach each other, the wave function overlaps and the off-diagonal elements of the Hamiltonian become non-negligible, and the energy eigenvalue gradually deviates from $H_{ii}$.
We calculate the coefficients for these $\sigma$ and $\pi$ orbits, as shown in Fig.~\ref{fig:coefficients}. When two binary black holes are separated by more than $30r_b$ away, the coefficients $f_{10}$ and $g_{11}$ approach $1$. At this point,  $\ket{\sigma_s}\approx\ket{\psi^1_{2p_x}}$ and $\ket{\pi_s}\approx\ket{\psi^1_{2p_y}}$, indicating that molecular orbits have not yet been formed. As the two binary black holes move closer, other coefficients become significantly nonzero, indicating formation of molecular orbitals.

\subsection{State evolution}

We now study the time evolution of scalar boson distribution in the binary black hole system as the companion black hole approaches to the primary. To simplify the analysis, we assume that initially the scalar bosons occupy exclusively the orbit $\ket{\psi^1_{211}}$ of the primary black hole, while the companion black hole is initially devoid of bosons. The situation in which bosons initially occupy the orbits of both black holes can be analyzed similarly. At $t \to -\infty$, the state of the boson can be formally expressed as a superposition of molecular orbitals, 
\begin{eqnarray}
    \ket{\Psi(t=-\infty)}\equiv \ket{\psi^1_{211}}
    =\sum_{a=1}^{2N} F_a \ket{\sigma^a_s}+\sum_{\mu=1}^N G_\mu \ket{\pi^\mu_s},
\end{eqnarray}
where
\begin{eqnarray}\label{eq:initial condition}
    F_a&=&\frac{(-1)^{a+1} \mathbf{D}_a}{\sqrt{2}\sum_{b=1}^{2N}(-1)^b f_{b0}\mathbf{D}_b},\nonumber\\
    G_\mu&=&\frac{i (-1)^{\mu} \mathbf{M_\mu}}{\sqrt{2}\sum_{\nu=1}^N (-1)^{\nu+1}g_{\nu 1}\mathbf{M_\nu}}.
\end{eqnarray}
We denote $\mathbf{D}_a = \text{det}(D_{-a})$, where $D_{-a}$ is a matrix obtained by removing the $a$-th row from the matrix $D$ given by  
\begin{eqnarray}\label{eq:D_matrix}
D=
\begin{pmatrix}
f_{11} &  \cdots  & f_{1(N+1)} & \cdots & f_{1(2N-1)} \\
\vdots & \ddots & \vdots & \ddots & \vdots & \\
f_{a1} & \cdots  & f_{a(N+1)} & \cdots & f_{a(2N-1)} \\
\vdots & \ddots & \vdots & \ddots & \vdots & \\
f_{(2N)1} & \cdots  & f_{(2N)(N+1)} & \cdots & f_{(2N)(2N-1)}
   \end{pmatrix};
\end{eqnarray}
and $\mathbf{M}_{\mu}=\text{det}(M_{-\mu})$, where $M_{-\mu}$ is a matrix obtained by removing the $\mu$-th row from the matrix $M$ given by 
\begin{eqnarray}
M= \begin{pmatrix}
g_{12} & g_{13} & \cdots &  g_{1N} \\
\vdots & \vdots & \ddots & \vdots \\
g_{\mu 2} & g_{\mu 3} & \cdots & g_{\mu N} \\
\vdots & \vdots & \ddots & \vdots \\
g_{N 2} & g_{N 3} & \cdots & g_{NN}
   \end{pmatrix}.
\end{eqnarray}

Since the orbit of the binary black holes shrinks extremely slow due to the emission of gravitational waves, the time evolution of the state of the bosons can be well approximated by an instantaneous Schr\"odinger equation. This implies that the wave functions of molecular orbitals are time dependent; however, they still form a complete set of bases. The state of the boson can be expanded using the molecular orbitals at any given time $t$, namely, 
\begin{eqnarray}
\label{eq:scalar intial condition}
    \ket{\Psi(t)}&=&\sum_{a=1}^{2N} F_a(t) e^{-i\int_{-\infty}^t E_\sigma^a d\tau } \ket{\sigma^a_s(t)}\nonumber\\
    &&+\sum_{\mu=1}^N G_\mu(t)e^{-i\int_{-\infty}^t E_\pi^\mu d \tau}\ket{\pi_s^\mu (t)}.
\end{eqnarray}
The time evolution consists of two parts: the exponential factor represents the free time evolution, and the coefficients $F_a$ and $G_\mu$ characterize the degree of participation of each molecular orbital. 
The time-dependent coefficients indicate that bosons may transfer from the primary black hole to the companion. The amount of scalar bosons that transfers to the companion can be evaluated by finding the coefficients of the atomic orbitals belonging to the companion. In particular, if the bosons transfer to the decaying modes $\ket{\psi^2_{ns}}$ and $\ket{\psi^2_{n1{-}1}}$, they could be absorbed by the companion black hole, resulting in boson cloud depletion. Since the atomic orbital $\ket{\psi^2_{ns}}$ has non-zero overlap only with the molecular orbitals $\ket{\sigma_s^a}$, so its corresponding coefficient in the state $\ket{\Psi(t)}$ can be found as
\begin{eqnarray}\label{eq:decay coef ns}
    C_{ns}(t)=\sum_{a=1}^{2N} F_a(t) e^{-i\int_{-\infty}^t E_{\sigma}^a d\tau}f_{an}(t).
\end{eqnarray}
Note that the time dependence of $f_{a n} (t)$ originates from the time-dependent molecular orbitals. The atomic orbital $\ket{\psi^2_{n1{-}1}}$ overlaps with both the molecular orbitals $\ket{\sigma_s^a}$ and $\ket{\pi_s^\mu}$, so the corresponding coefficient is given by 
\begin{eqnarray}\label{eq:decay coef n11}
    C_{n1-1}(t)=& \frac{1}{\sqrt{2}} \sum_{a=1}^{2N} F_a(t) e^{-i\int_{-\infty}^t E^a_\sigma d\tau}f_{a(N+n-1)}(t)\nonumber\\
    & + \frac{i}{\sqrt{2}} \sum_{\mu=1}^N G_\mu(t)e^{-i\int_{-\infty}^t E^\mu_\pi d\tau}g_{\mu n}(t).
\end{eqnarray}
The squared modulus of these coefficients represents the probability of boson transfer to the atomic orbital $\ket{\psi^2_{ns}}$,
\begin{eqnarray}
       |C_{ns}(t)|^2&=&\sum_{a=1}^{2N} |F_a(t)|^2 f_{an}(t)^2+\sum_{a\neq b}F_a^*(t) F_b(t)\nonumber\\
       &&\times f_{an}(t)f_{bn}(t)\cos\int_{-\infty}^t(E^a_\sigma-E^b_\sigma)d\tau,
\end{eqnarray}
and to the atomic orbital $\ket{\psi^2_{n1-1}}$,
\begin{eqnarray}\label{eq:scalar occupation density p}
        &&|C_{n1-1}(t)|^2=\nonumber\\
         &&\frac{1}{2}\bigg\{\sum_{a=1}^{2N} |F_a(t)|^2 f_{a(N+n-1)}(t)^2 +\sum_{\mu=1}^N  |G_\mu(t)|^2 g_{\mu n}(t)^2 \nonumber\\
        &&+\sum_{a\neq b}F_a^*(t) F_b(t)f_{a(N+n-1)} f_{b(N+n-1)}\nonumber\\
        &&\times\cos\int_{-\infty}^t(E^a_\sigma-E^b_\sigma)d\tau
         + i\sum_{a,\mu}\Big[F_a^*(t) G_\mu(t)\nonumber\\
         &&-G_\mu^*(t) F_a(t)\Big]
         \times f_{a(N+n-1)}g_{\mu n} \cos\int_{-\infty}^t(E^a_\sigma-E^\mu_\pi)d\tau \nonumber\\
        && +\sum_{\mu\neq \nu}G_\mu^*(t) G_\nu(t)
        g_{\mu n}g_{\nu n} \cos\int_{-\infty}^t(E^\mu_\pi-E^\nu_\pi)d\tau \bigg\}.
\end{eqnarray}

Here, $F_a (t)$ and $G_\mu (t)$ represent the coefficients of the $a$-th and $\mu$-th molecular orbitals, respectively, with their initial values given by Eq.~\eqref{eq:initial condition}. 
The squared modulus $|F_a(t)|^2$ represents the probability that the boson occupies the $a$-th molecular orbital, while $|G_\mu(t)|^2$ represents the probability of occupying the $\mu$-th molecular orbital. The lowercase letters $f_{ai}$ and $g_{\mu i}$ denote the contributions of the $i$-th atomic orbital to the $a$-th $\sigma$ molecular orbital and the $i$-th atomic orbital to the $\mu$-th $\pi$ molecular orbital, respectively. For example, the squared modulus $|f_{a(N+n-1)}|^2$ represents the probability of occupying the mode $\ket{\psi^2_{n p_x}}$ state, given that the boson is in the $a$-th $\sigma$ molecular orbital. The similar interpretation applies to other $f$ and $g$ coefficients. The coefficients $f$ and $g$ can be calculated from Eq.~\eqref{eq: solve_coefficients}, whereas the energy eigenvalues are determined using Eq.~\eqref{eq:determinant}. The only remaining task is to solve for the time-dependent coefficients $F_a (t)$ and $G_\mu (t)$.

Since the time-dependent parameter, the orbital separation between two black holes, in the Hamiltonian varies very slowly, one might expect that the adiabatic approximation applies. This implies that bosons remain in their initial molecular orbit and cannot transition to other molecular orbits. However, after a very long evolution, there might be a small fraction of bosons transitioning to other molecular orbits. We take into account this non-adiabatic evolution and estimate its effect on the boson transfer.

Consider a generic quantum state 
\begin{eqnarray}\label{eq:generic-state}
\ket{\psi(t)}=\sum_a c_a(t)\ket{\psi_a(t)},
\end{eqnarray}
which is expanded in terms of a set of orthonormal and complete instantaneous energy eigenstates $\{ \ket{\psi_a(t)} \}$, with $c_a(t)$ the coefficients. If the adiabatic approximation is valid, then we have $|c_a(t+dt)|= |c_a(t)|$, showing that there is no transition between different energy levels. By substituting the state $\ket{\psi(t)}$ into the instantaneous Schr\"odinger equation, we obtain the time derivative of the coefficient $c_a(t)$,
\begin{eqnarray}\label{eq:adiabatic approximation}
 \frac{\partial c_a}{\partial t}=-i c_a E_a -\sum_{b \neq a} c_b \frac{1}{E_b - E_a} \bigg\langle \psi_a \bigg| \frac{\partial \hat{H}}{\partial t} \bigg| \psi_b \bigg\rangle,
\end{eqnarray}
where the first term characterizes the free time evolution of the initial energy eigenstate, while the second term could induce transitions from the initial energy eigenstate into other energy eigenstates. It is evident that the adiabatic approximation applies if the sum in the second term vanishes, which turns out to be the case for bosons in a binary black hole system with equal masses. 

However, the adiabatic approximation does not generally hold for a binary black hole system with unequal masses. Although the amplitude of the transition term is small compared to the free time evolution term, the contribution of the transition term accumulates gradually and becomes significant after a long-time evolution.

We separate the free time evolution part from the coefficient $c_a(t)$ and define $G_a(t)$ as
\begin{eqnarray}
    c_a(t) = G_a(t) e^{-i\int_{-\infty}^t E_a d\tau}.
\end{eqnarray}
By substituting $c_a(t)$ into Eq.~\eqref{eq:adiabatic approximation} one can obtain the time derivative of $G_a(t)$,
\begin{eqnarray}\label{eq:non-adiabatic evolution}
    \frac{\partial G_a}{\partial t}
    =-\sum_{b\neq a}\frac{G_b e^{-i\int_{-\infty}^t(E_b-E_a)d\tau}}{E_b-E_a} \bigg\langle \psi_a \bigg| \frac{\partial \hat{H}}{\partial t} \bigg| \psi_b \bigg\rangle,
\end{eqnarray}
from which one can numerically solve for the coefficient $G_a(t)$.

In the equal-mass case, two bonding molecular orbitals and two anti-bonding molecular orbitals exist even when the binary black holes are infinitely far apart. This arises because the two black holes have identical masses, ensuring that the energy levels $\ket{\psi^1_{2p_x}}$ and $\ket{\psi^1_{2p_y}}$ of the primary black hole, used in the construction of molecular orbitals, are degenerate with $\ket{\psi^2_{2p_x}}$ and $\ket{\psi^2_{2p_y}}$ of the companion black hole at any separation. Consequently, the coefficients of the atomic orbitals always have the same magnitude, differing only in sign for anti-bonding molecular orbitals. Since we assume that the boson cloud occupies one of the atomic orbitals of the primary black hole when the orbital separation $R$ is sufficiently large, 
the probability of finding the boson in the $\ket{\psi^2_{21-1}}$ state of the companion  approaches zero~\cite{guo2023masstransferbosoncloud}. 

On the other hand, if the two black holes have unequal masses, the energy levels $\ket{\psi^1_{2p_x}}$ and $\ket{\psi^1_{2p_y}}$ of the primary black hole are not degenerate with the corresponding states of the companion. Consequently, for a given molecular orbital labeled with $a$, the coefficients $f_{ai}$ are not equal for different $i$ at any separation. This is also the case for the coefficients $g_{\mu j}$. In the limit of infinite separation, the molecular orbitals reduce to the atomic orbitals of either the primary or companion black hole. Consider the $\sigma$ molecular orbitals as an example. For $a = 1$, we find $f_{10} = 1$ and $f_{11} = f_{12}=\cdots = 0$, which implies that the molecular orbital $\sigma^1_s$ reduces to the $\ket{\psi^1_{2 p_x}}$ state of the primary black hole. In contrast, for $a = 2$ and higher, $f_{a0} = 0$, which implies that the corresponding molecular orbitals $\sigma^2_s$ and others reduce to atomic orbitals of the companion black hole. 
Since we assume that the boson cloud is initially located around the primary black hole, $|C_{ns}|^2$ and $|C_{n1-1}|^2$ approach zero when the binary black holes are infinitely far apart, which can be consistently checked as follows.
According to the initial condition specified by Eq.~\eqref{eq:initial condition}, the only nonzero coefficients are $F_1$ and $G_1$, and $F_2 = F_3 = G_2 =\cdots = 0$. While for $a =1$, we have $f_{11} = f_{12} = g_{12} =\cdots = 0$; therefore, the summations $\sum_{a=1}^{2N}|F_a|f_{an}$, $\sum_{a=1}^{2N}|F_a|f_{a(N+n-1)}=0$ and $\sum_{a\neq b}F_{a}F_{b}^*$ vanish. Consequently, both $|C_{ns}|^2$ and $|C_{n1-1}|^2$ approach zero.

As binary black holes move closer, the gravitational field of the companion black hole alters the Hamiltonian of the initial BH-boson system, inducing changes in both the eigenvalues and eigenvectors of molecular orbitals. In the case of equal mass, a nonzero probability of occupying the atomic orbitals of the companion was observed~\cite{guo2023masstransferbosoncloud}. In the unequal-mass case, the molecular eigenvectors, characterized by the coefficients $f_{ai}$ and $g_{\mu j}$, are also modified. Some coefficients that were initially zero--such as $f_{11}$, $f_{12}$,$\cdots$, $f_{1(2N)}$, as well as $f_{20}$, $f_{30}$, $\cdots$, become nonzero, leading to the formation of molecular orbitals. Furthermore, because of the non-adiabatic evolution, bosons may transfer between different molecular orbitals. The initially vanishing coefficients $F_a$ and $G_\mu$ also acquire non-zero values. Consequently, the occupation probabilities $|C_{ns}(t)|^2$ and $|C_{n1-1}(t)|^2$ gradually increase, showing that bosons can transfer to the companion black hole and occupy decaying modes $\ket{\psi^2_{ns}}$ and $\ket{\psi^2_{n1-1}}$.

\begin{figure}[htbp]
    \centering    \includegraphics[width=1\columnwidth]{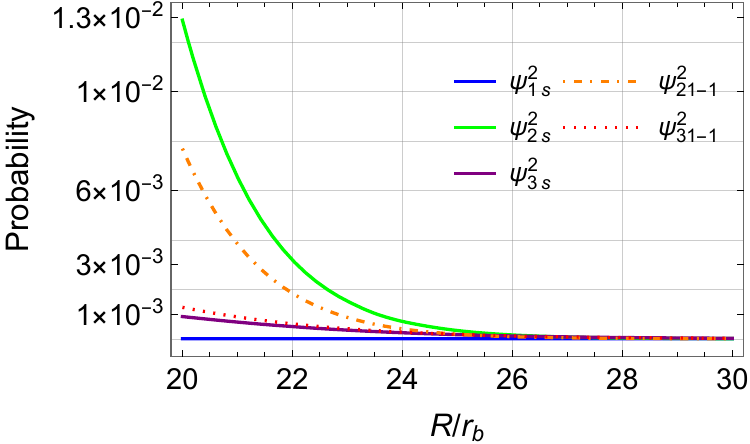}
    \caption{Occupation probabilities of the companion black hole's decaying modes, evaluated for $q=0.9$ and $\alpha=0.1$.}
    \label{fig:occupation probability}
\end{figure}

Figure~\ref{fig:occupation probability} shows the occupation probabilities $|C_{ns}(t)|^2$ and $|C_{n1-1}(t)|^2$ for the case with parameters $q=0.9$ and $\alpha=0.1$. In the numerical calculation, we take a sufficiently large initial orbital separation $R_*=80r_b$ as a proxy for the infinitely large distance ($t=-\infty$) in the lower limit of integration. It can be seen that when the binary black holes are far apart, no effective molecular orbitals are formed, and thus both probabilities approach zero. As the binary separation decreases, however, the wave functions begin to overlap effectively, and the corresponding occupation probabilities gradually increase. The behavior is qualitatively similar for other choices of parameters.

Bosons occupying the decaying modes of the companion black hole could be absorbed by it, leading to the depletion of the boson cloud. In the following analysis, we neglect both the back reaction of boson transfer on the binary orbital evolution and its effect on the decay rate. There are two types of cloud depletion channels, one through the decaying modes $\ket{\psi^2_{ns}}$ and the other through the decaying modes $\ket{\psi^2_{n,1,-1}}$. 
Therefore, the rate of change of the total mass $M_c$ of the boson cloud can be expressed as 
\begin{eqnarray}\label{eq:total depletion}
\frac{dM_c}{dt}=2\sum_{n}(|C_{n1-1}|^2\Gamma_{n1-1}+|C_{ns}|^2\Gamma_{n00})M_c.
\end{eqnarray}

\begin{figure}[htbp]
    \centering
    \includegraphics[width=1\columnwidth]{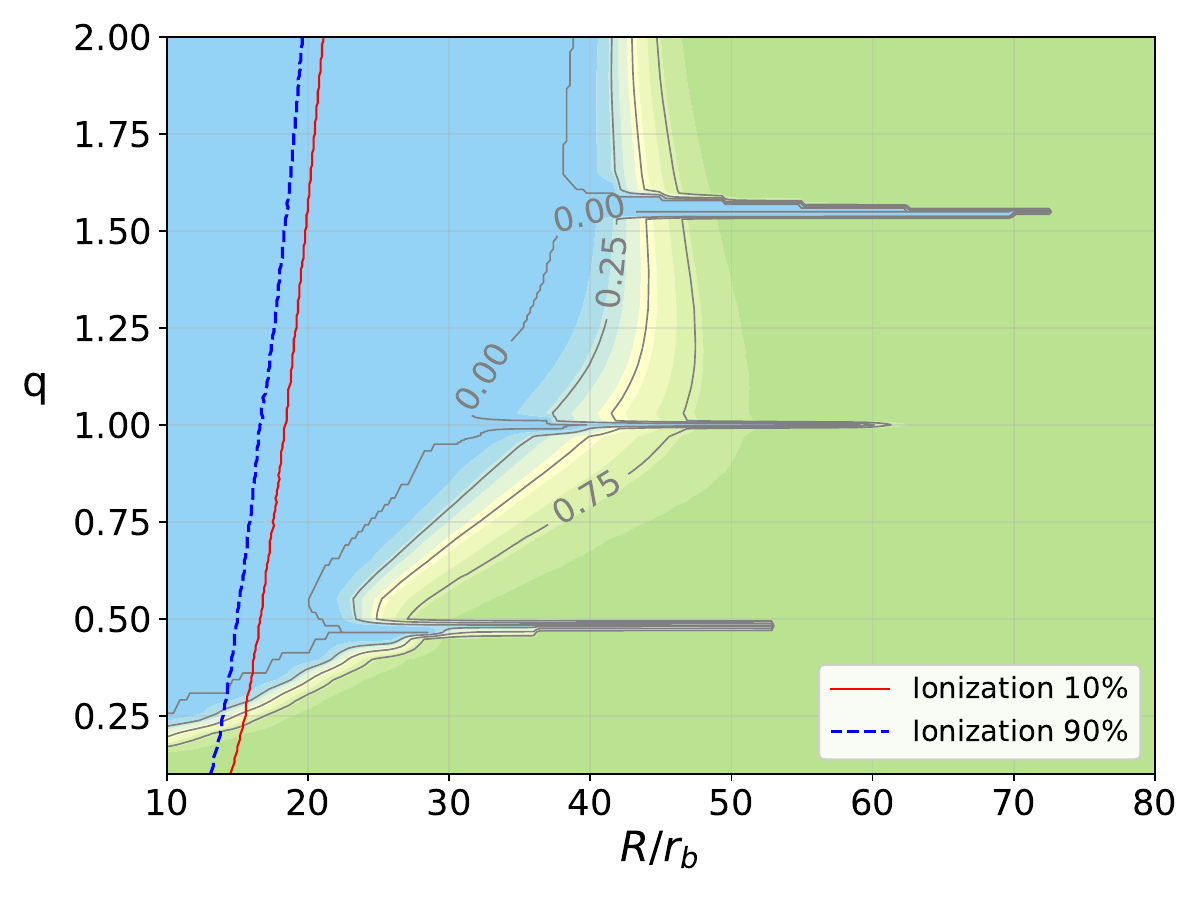}
    \caption{Scalar cloud depletion via molecular orbitals and dominant separations of ionization. The green region indicates no cloud depletion, the blue region represents sharp cloud depletion, and the red curve corresponds to ionization removing approximately $10\%$ of the cloud, another blue curve represent $90\%$ ionization of the cloud. Here, we use the fine-structure-like constant $\alpha=0.1$.
    }
    \label{fig:scalar total depletion}
\end{figure}

To illustrate the characteristics of cloud depletion, we calculate $M_c/M_0$, the ratio between the cloud mass $M_c$ at any time and the initial cloud mass $M_0$, as shown in Fig.~\ref{fig:scalar total depletion} for $\alpha = 0.1$. It can be seen that the boson cloud starts to deplete and is completely absorbed at orbital separations where the probability of occupying the companion black hole's decaying modes remains small. This occurs because the shrinkage of the binary orbit is rather slow due to the emission of gravitational waves, leading to a very long evolutionary timescale that significantly enhances the cloud depletion, even though the occupation probability is small.

We consider boson cloud depletion in a binary black hole system with different mass ratios. The case of equal mass, i.e., $q = 1$, has been studied in detail in Ref.~\cite{guo2023masstransferbosoncloud}. We evaluated $M_c/M_0$ for cases of unequal masses, i.e., $q \ne 1$, as shown in Fig.~\ref{fig:scalar total depletion} for $\alpha = 0.1$. To focus exclusively on the cloud depletion caused by boson transfer, we first neglect other dynamical effects, e.g., hyperfine and Bohr mixing. Furthermore, we assume that the boson cloud exists at orbital separations $R \ge 80 r_b$. Contributions to cloud depletion through decaying modes with different principal quantum numbers $n$ depend on the mass ratio. In most cases, the contribution from a single quantum number dominates. However, for certain mass ratios, contributions from two or more quantum numbers become comparable. Therefore, we consider principal quantum numbers ranging from $n=1$ to $n=5$, which cover the dominant contributions for the mass ratios examined in this work.

It is evident from Fig.~\ref{fig:scalar total depletion} that, for sufficiently large mass ratios, the boson cloud depletes completely at a particular orbital separation, showing that the mass transfer and the subsequent depletion of the boson cloud are important parts of boson cloud dynamics in a binary black hole system. 
There exist certain mass ratios that yield significantly larger depletion distances than others. These characteristic mass ratios are approximately multiples of $1/2$. At infinite separation between the two black holes, the energy of the companion's atomic orbital with principal quantum number $n$, relative to the primary's $\ket{\psi_{2p_x}^1}$ orbital, scales as $2q/n$. Therefore, when the mass ratio $q$ approximately matches $n/2$, the energy levels of the two black holes coincide, allowing molecular orbitals to form at larger separations. 
A detailed discussion of these special mass ratios is provided in the Appendix~\ref{appendix special mass ratio}. When the mass ratio is very small, the cloud cannot be fully depleted into the companion. This occurs because the energy eigenvalue of the companion's lowest state, $\ket{\psi^2_{1s}}$, differs substantially from that of the primary state, $\ket{\psi^1_{2p_x}}$, preventing the formation of effective molecular orbitals. This indicates that a small companion cannot efficiently capture bosons from the primary.

\subsection{Cloud depletion due to other dynamics}

Ionization occurs when the companion black hole is sufficiently close to the primary black hole. Generally, the orbital separation at which ionization begins for the counter-rotating orbits is larger than that for the co-rotating orbits. The ionized bosons escape from the gravitational potential of the binary system, leading to the depletion of the boson cloud. After a sufficiently long time, the cloud could be completely dispersed through ionization.

\begin{figure}[htbp]
    \centering
    \includegraphics[width=1\columnwidth]{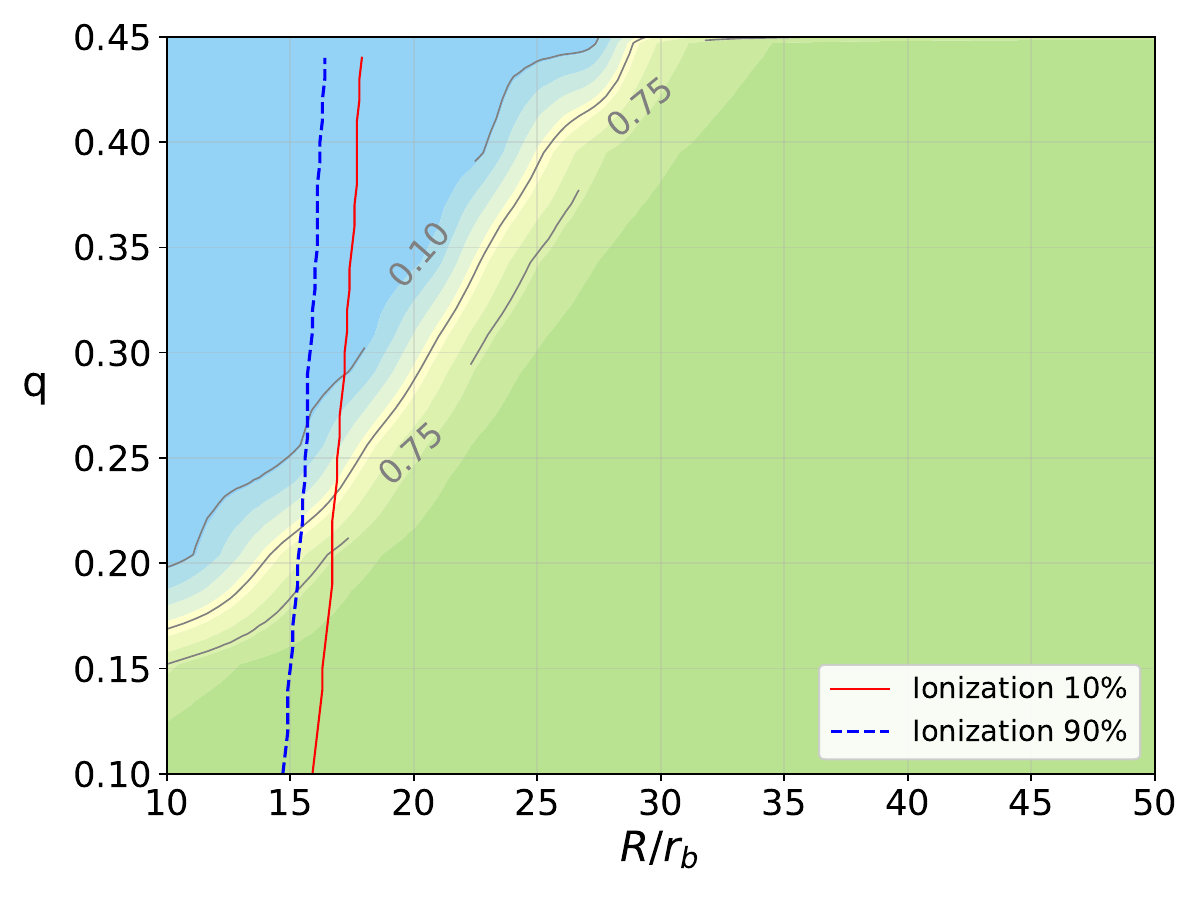}
    \caption{Scalar cloud depletion by molecular orbitals and the dominant separation of ionization. The green region indicates no cloud depletion, the blue region represents sharp cloud depletion, and the red curve corresponds to ionization removing approximately $10\%$ of the cloud, another blue curve represent $90\%$ ionization of the cloud. Here, we choose the fine-structure-like constant $\alpha=0.05$.
    }
    \label{fig:scalar_ionization_0.05}
\end{figure}

We explicitly calculate the ratio $M_c/M_0$ as a function of the orbital separation $R_*$, considering the effect only due to ionization for counter-rotating orbits, as shown in Fig.~\ref{fig:scalar total depletion} for $\alpha = 0.1$. It is evident that for sufficiently large mass ratios, the typical orbital separations at which cloud depletion occurs through mass transfer are larger than those for ionization. This indicates that the mass transfer to the companion black hole plays a dominant role in the cloud dynamics. 
However, for small mass ratios, the formation of molecular orbitals is inefficient, and only a small fraction of the boson cloud can be depleted through mass transfer. In this regime, the effect of ionization becomes dominant. We further vary the parameter $\alpha$ and find a similar trend: mass transfer dominates at large mass ratios, whereas ionization dominates at small mass ratios, as shown for $\alpha=0.05$ in Fig.~\ref{fig:scalar_ionization_0.05}.

The gravitational perturbation from the companion can induce mixing of boson's energy levels around the primary black hole, e.g., hyperfine and Bohr mixings, and subsequently lead to boson cloud depletion~\cite{PhysRevD.99.044001}. 
For co-rotating orbits, hyperfine mixing occurs at much larger orbital separations than mass transfer, causing most of the cloud to be depleted into the primary black hole. For counter-rotating orbits, although the hyperfine resonance does not occur, weak hyperfine mixing can still lead to substantial cloud depletion over the long inspiral timescales. For both co-rotating and counter-rotating orbits, only a small fraction of the boson cloud remains when mass transfer begins.

\begin{figure}[htbp]
    \centering
    \includegraphics[width=1\columnwidth]{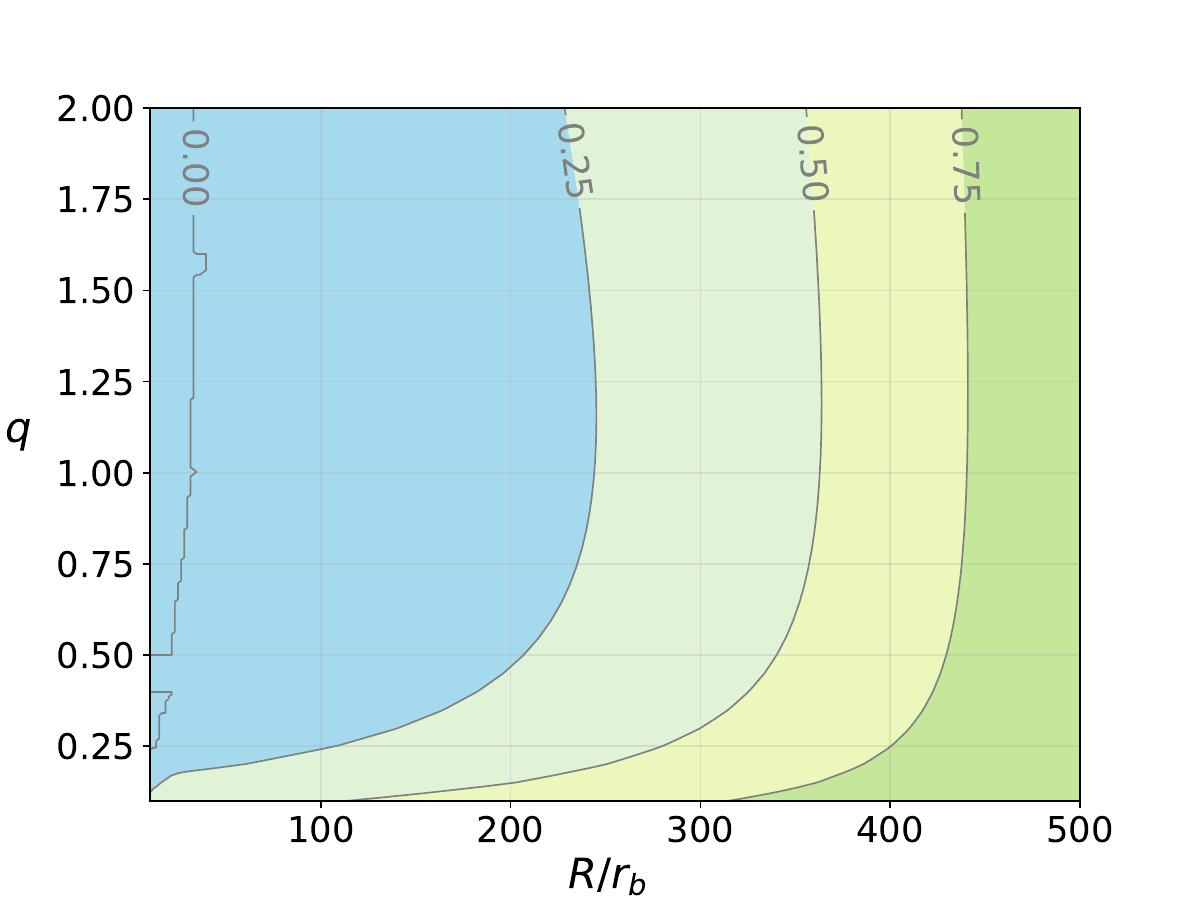}
    \caption{Scalar cloud depletion in counter-rotating orbits, incorporating hyperfine mixing, Bohr mixing and mass transfer, with the fine-structure-like constant $\alpha=0.1$.
    }
    \label{fig:scalar hyper bohr depletion}
\end{figure}

We calculate the ratio $M_c/M_0$ as a function of the orbital separation $R_*$ for various mass ratios, taking into account the effects of
hyperfine mixing, Bohr mixing, and mass transfer, as shown in Fig.~\ref{fig:scalar hyper bohr depletion} for $\alpha = 0.1$. To highlight the significant role of hyperfine mixing, we assume that the primary black hole hosts a boson cloud at orbital separation $R_* = 500r_b$, while the companion does not host a boson cloud. It is evident from Fig.~\ref{fig:scalar hyper bohr depletion} that most of the boson cloud has been absorbed by the primary black hole through hyperfine mixing when the dynamics begins around $R_* = 500 r_b$, indicating that hyperfine mixing plays a dominant role in cloud dynamics. 

\begin{figure}[htbp]
    \centering
    \includegraphics[width=1\columnwidth]{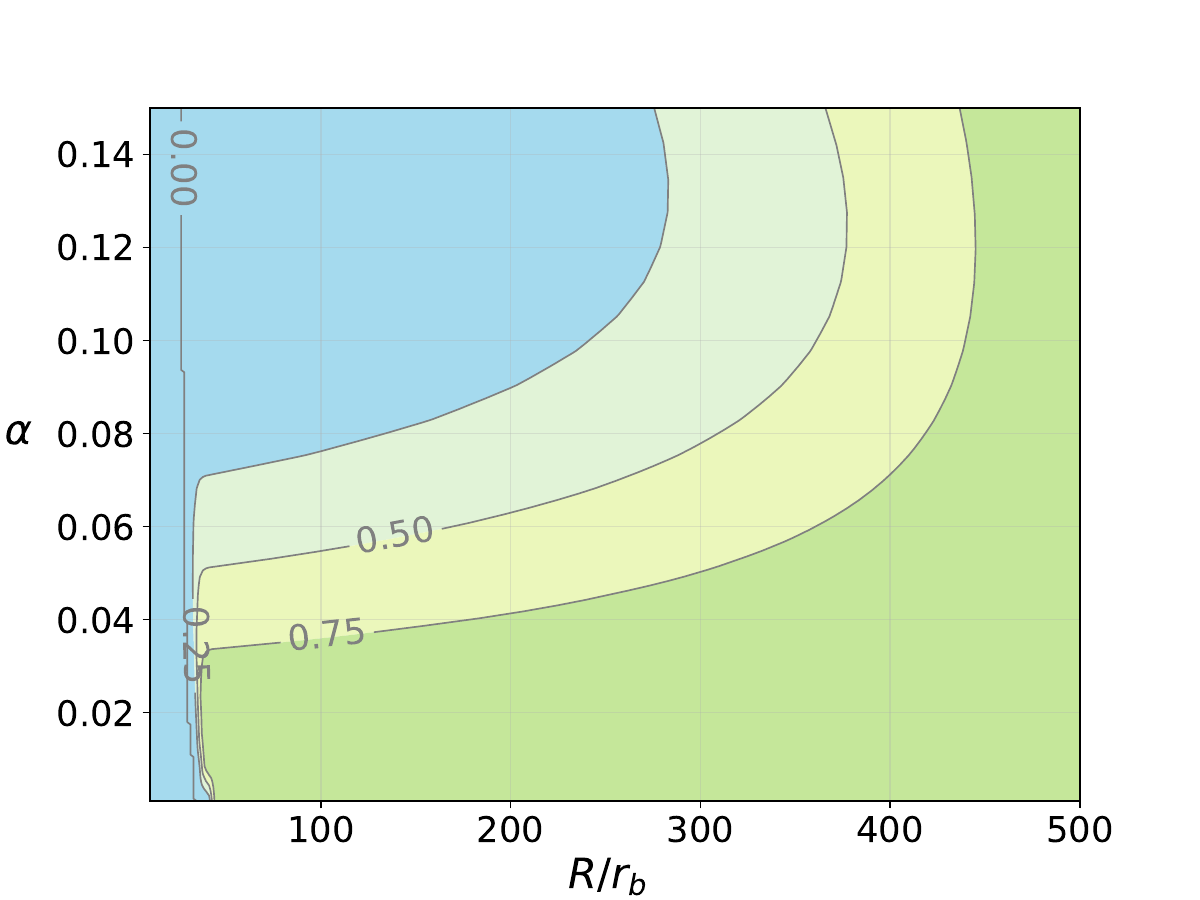}
    \caption{Scalar cloud depletion in counter-rotating orbits, involving hyperfine mixing, Bohr mixing and mass transfer, for a mass ratio $q=0.9$. The cloud is initialized at an orbital separation of $500r_b$.
    }
    \label{fig:scalar alpha dependence}
\end{figure}

We further calculate the ratio $M_c/M_0$ as a function of orbital separation $R_*$ for various $\alpha$, taking into account the effects of
hyperfine mixing, Bohr mixing, and mass transfer, as shown in Fig.~\ref{fig:scalar alpha dependence} for $q = 0.9$. 
In this analysis, 
the boson cloud is assumed to form around the primary when the orbital separation is approximately $R_*=500r_b$. 
It can be seen that when $\alpha \ge 0.07$ the boson cloud suffers from significant depletion at larger separations compared to those with smaller $\alpha$. In this regime, the cloud depletion is dominated by perturbative resonance and level mixing, and the boson cloud is primarily absorbed by the primary black hole. In contrast, for $\alpha \le 0.07$, the decay rates of the atomic orbitals to which the boson is transferred via resonance and level mixing are very low, and most of the cloud depletes at separations where the molecular orbitals have already formed. Consequently, the majority of the bosons are absorbed by the companion black hole.

\section{Vector cloud}
\label{Sec: Vector_cloud}
In this section, we investigate the dynamics of vector bosons in a binary black hole system, focusing particularly on mass transfer to the companion black hole and the resulting cloud depletion. Since cloud depletion due to hyperfine and Bohr mixings is comparatively weak and negligible, mass transfer could dominate the cloud depletion. 

\subsection{Molecular orbitals for vector bosons}

We denote by $\ket{nljm}$ or $\ket{\psi_{nljm}}$ the state of a vector boson, with $n$ the principal quantum number, $l$ the orbital angular momentum, $j$ the total angular momentum, and $m$ the azimuthal quantum number.
Since the dominant growing mode of the vector boson generated through superradiance is the $\ket{1011}$ mode, we therefore assume that the primary black hole is initially surrounded by a vector boson cloud, with all bosons occupying the state $\ket{\psi^1_{1011}}$. For simplicity, we further assume that the companion black hole is initially empty of bosons. 
Similar to the scalar case, vector bosons can also transfer from the primary black hole to the companion black hole during the evolution of the binary system. The Hamiltonian that governs the dynamics of the vector boson follows Eq.~\eqref{eq:Hamiltonian}, which exhibits rotational symmetry about the $x$-axis. Consequently, transitions to the companion are restricted to the $s$ and $p$ states, corresponding to the orbital angular momentum quantum numbers $l=0$ and $l=1$, respectively. The states with $l=0$ are referred to as $s$ states and possess spherical symmetric spatial distributions. However, due to the presence of spin, an orbital angular momentum of $l=1$ can correspond to total angular momenta $j=0,1,2$. This implies that the ways in which $p$-orbitals can be combined are distinct. In the following, we focus on the construction of the $p$ state of the vector boson. 

\begin{figure}[htbp]
    \centering
    \includegraphics[width=1\linewidth]{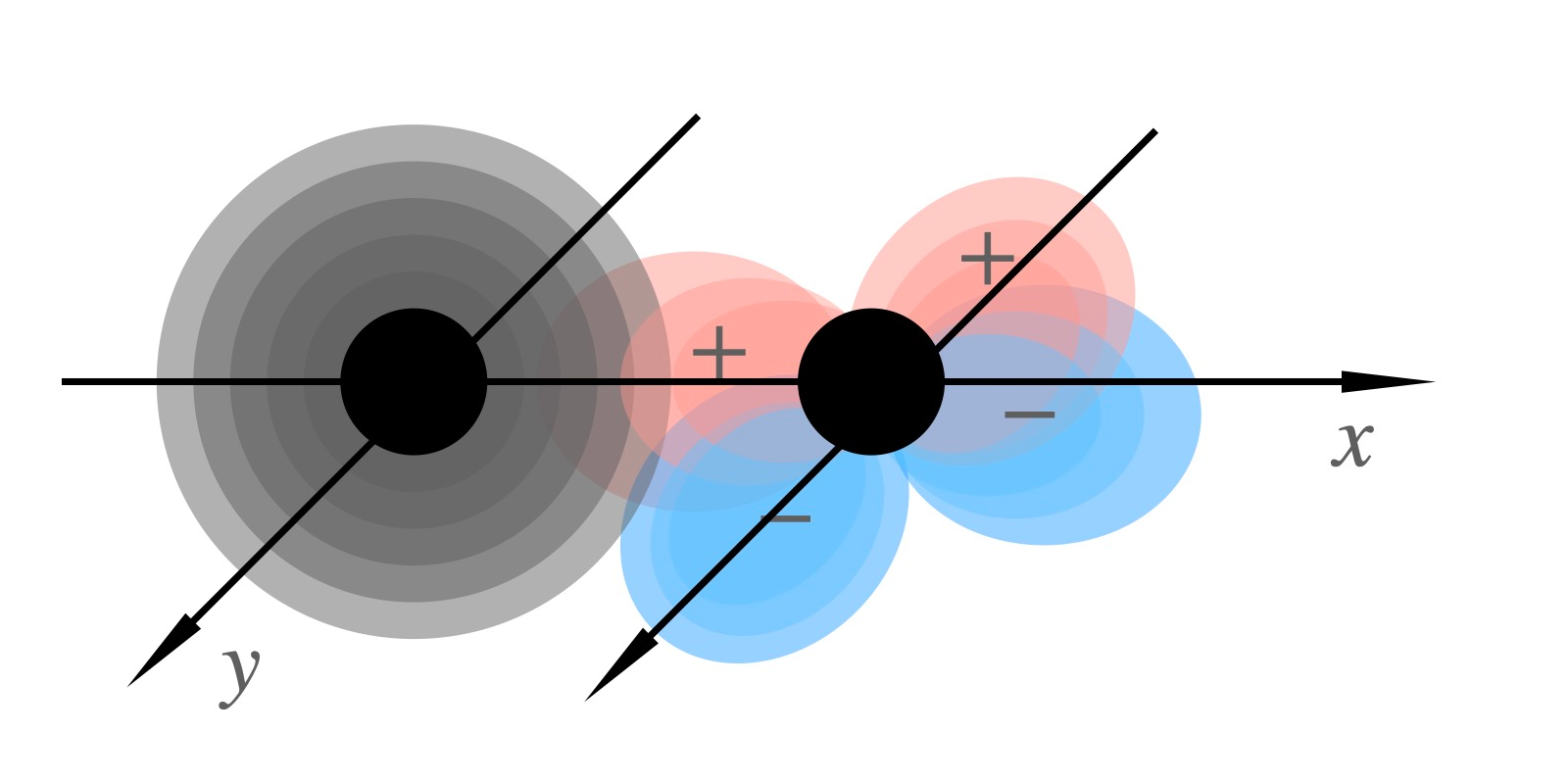}
    \caption{Illustration of a binary black hole system: the central black hole is surrounded by a vector boson cloud in the spherically symmetric state $\ket{1011}$. The companion black hole depletes the boson cloud via its $p$-state. The pink regions represent wavefunction components with a positive sign, while the blue regions represent negative components. As the companion black hole approaches, the $p_y$ state overlaps with the $\ket{1011}$ state with both positive and negative components, leading to a cancellation.}
    \label{fig:binary black holes}
\end{figure}

In contrast to the scalar case, the intrinsic spin of vector bosons enlarges the state space and complicates the construction of $p$ states. 
We denote by $\ket{\mathbf{\xi}^{m_s}\psi_{nlm}}$ the state with a definite spin orientation, where $m_s = +1, 0, -1$ correspond to spin aligned, perpendicular, and anti-aligned with respect to the black hole's $z$ axis, respectively. Since wave functions with different spin orientations are mutually orthogonal, vector bosons in the atomic orbitals with a definite spin projection $m_s$ can only overlap effectively with the companion's atomic orbitals carrying the same spin projection. 
Therefore, it is necessary to transform the spin–orbit–coupled basis into one in which the spin and orbital angular momentum are treated separately. For orbital angular momentum $l=1$, coupling with spin angular momentum $s=1$ yields total angular momentum states with $j=0,1,2$. The relation between the spin–orbit–coupled basis states and the basis states with separated orbital and spin angular momenta can be expressed using the Clebsch–Gordan coefficients~\cite{PhysRevD.96.035019}.
\begin{eqnarray}\label{eq:basisTrans}
    \ket{\psi_{n100}}&=&\frac{1}{\sqrt{3}} \big( \ket{\mathbf{\xi}^{+1}\psi_{n1-1}}-\ket{\xi^0 \psi_{n10}}
    +\ket{\xi^{-1}\psi_{n11}} \big), \nonumber\\
    \ket{\psi_{n110}}&=&\frac{1}{\sqrt{2}} \big( \ket{\xi^{+1}\psi_{n1-1}}-\ket{\xi^{-1} \psi_{n11}} \big), \nonumber\\
    \ket{\psi_{n120}}&=&
    \frac{1}{\sqrt{6}} \big( \ket{\xi^{+1}\psi_{n1-1}}+2\ket{\xi^0 \psi_{n10}} 
    +\ket{\xi^{-1}\psi_{n11}} \big), \nonumber\\
\end{eqnarray}
where $n$ denotes the principal quantum number, and the superscript of $\xi$ specifies the spin orientation. Since the atomic state $\ket{\psi^1_{1011}}$ of the primary has a spin projection $m_s=+1$, only the companion states with $m_s=+1$ have nonzero overlap with $\ket{\psi^1_{1011}}$, and thus contribute to the formation of molecular orbitals. By inverse the basis transformation in Eq.~\eqref{eq:basisTrans}, we find
\begin{eqnarray}\label{eq:spin-state}
   \ket{\xi^{+1}\psi_{n1-1}} &=& \frac{1}{\sqrt{3}}\ket{\psi_{n100}}+\frac{1}{\sqrt{2}}\ket{\psi_{n110}}+\frac{1}{\sqrt{6}}\ket{\psi_{n120}}, \nonumber\\
   \ket{\xi^{-1}\psi_{m11}}&=&\frac{1}{\sqrt{3}}\ket{\psi_{n100}}-\frac{1}{\sqrt{2}}\ket{\psi_{n110}}+\frac{1}{\sqrt{6}}\ket{\psi_{n120}}, \nonumber \\
   \ket{\xi^{0}\psi_{n10}}&=&-\frac{1}{\sqrt{3}}\ket{\psi_{n100}}+\frac{\sqrt{2}}{\sqrt{3}}\ket{\psi_{n120}}). 
\end{eqnarray}
Note that the wave functions of the states defined in Eq.~\eqref{eq:spin-state} are complex. To facilitate further analysis, it is convenient to construct states with definite spin orientations and real-valued wave functions. For instance, the states with spin orientation $m_s=1$ can be written as
\begin{eqnarray}
         \ket{\xi^{+1}\psi_{np_x}}&=&\frac{1}{\sqrt{2}}(\ket{\xi^{+1}\psi_{n1-1}}-\ket{\psi_{n122}}),\nonumber\\
        \ket{\xi^{+1}\psi_{np_y}}&=&\frac{i}{\sqrt{2}}(\ket{\xi^{+1}\psi_{n1-1}}+\ket{\xi^{+1}\psi_{n11}}).
\end{eqnarray}

It is found that the overlaps between the primary black hole's $\ket{\psi_{1011}^1}$ and the companion's $s$ or $p_x$ states increase from zero at infinite separation to finite values as the orbital separation decreases. In contrast, the overlap with the companion's $p_y$ state vanishes identically, because the spatial distribution of the combined wave function consists of two lobes located symmetrically on opposite sides of the $x$-$z$ plane, having equal magnitude but opposite sign. Consequently, the system supports only $\sigma$-type molecular orbitals, while $\pi$-type orbitals are absent. We define the $\sigma$ orbitals as
\begin{eqnarray}\label{eq:molecular-orbitals-general}
\ket{\sigma_v}&=&c_{a}\ket{\psi^1_{1011}}+\sum_{n=1}^N c_{ns}\ket{\psi^2_{n011}}\nonumber\\
&&+\sum_{n=2}^N c_{np_x}\ket{\xi^{+1}\psi_{np_x}^2},
\end{eqnarray}
where the subscript $v$ indicates that these molecular orbitals belong to vector bosons. Here, $N$ specifies the number of principal quantum numbers of the companion that are included in constructing the molecular orbitals: $N$ $s$-type orbitals and $N-1$ $p$-type orbitals, with the lowest principal quantum number for $p_x$ being $2$. 

The state $\ket{\sigma_v}$ has the same form as in the scalar case given by Eq.~\eqref{eq: scalar molecular orbits}, except that the basis states are replaced with $\{\psi^1_{1011}, \psi^2_{n011}, \xi^{+1}\psi^2_{np_x}\}$. The energy eigenvalues of the molecular orbitals $\sigma_v$ can be obtained from Eq.~\eqref{eq:determinant}, and the corresponding eigenvectors follow by substituting these eigenvalues into Eq.~\eqref{eq: solve_coefficients}.
The only modification arises in the matrix elements: $S_{ij}=\braket{\Psi_i\mid\Psi_j}$ and $H_{ij}=\bra{\Psi_i}H\ket{\Psi_j}$ are defined as before, but with the basis states $\Psi_i$ now taken from $\{\psi^1_{1011}, \psi^2_{n011}, \xi^{+1}\psi^2_{np_x}\}$. 

\begin{figure}[htbp]
    \centering
    \includegraphics[width=1\columnwidth]
    {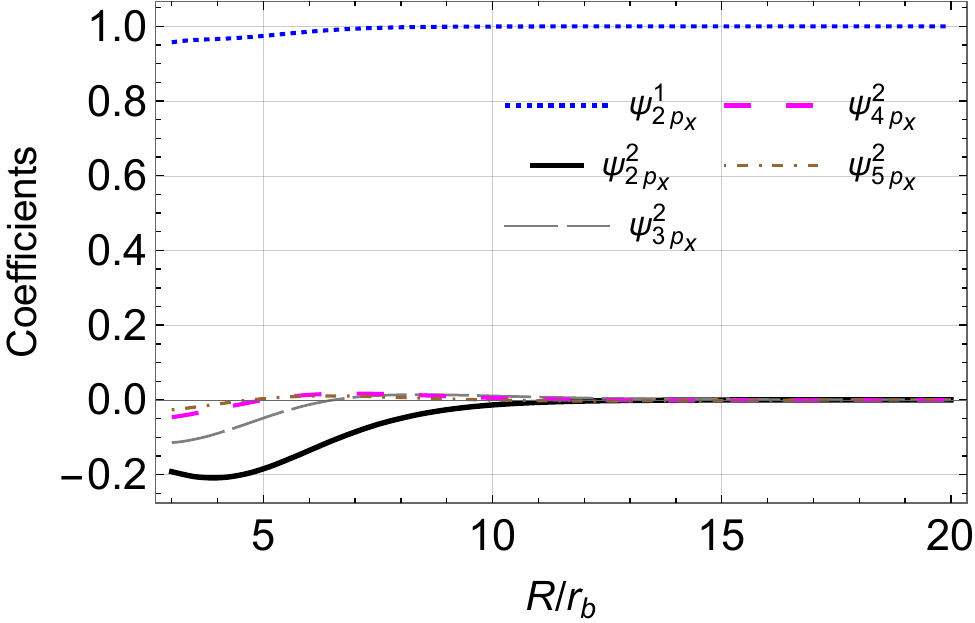}
    \caption{Coefficients of molecular orbitals for vector bosons corresponding to $E_0$. We choose $q=2$ and $\alpha=0.1$.
    }
    \label{fig:vector_coefficients}
\end{figure}
There are $2N$ basis states $\psi$ involved in Eq.~\eqref{eq:molecular-orbitals-general}, therefore, solving Eq.~\eqref{eq:determinant} yields $2N$ energy eigenvalues, denoted by $E_a$, with the corresponding eigenvectors forming the molecular orbitals:
\begin{eqnarray}
\ket{\sigma_v^a}&=&f_{a0}\ket{\psi^1_{1011}}+\sum_{n=1}^N f_{an}\ket{\psi^2_{n011}}\nonumber\\
    &&+\sum_{n=2}^N f_{a(N+n-1)}\ket{\xi^{+1}\psi_{np_x}^2}, 
\end{eqnarray}
where the symbol $a = 1, 2, \cdots, 2N$ indicates different molecular orbitals. Figure~\ref{fig:vector_coefficients} shows the eigenvectors of the molecular orbitals, respectively, with the mass ratio $q=2$ and the fine-structure-like constant $\alpha = 0.1$.

\subsection{Depletion of vector boson cloud}

We now study the dynamics of the vector boson cloud in a binary black hole system. To simplify the analysis, we assume that the vector bosons initially occupy only the $\ket{\psi^1_{1011}}$ state of the primary black hole, while the companion is initially devoid of bosons. A scenario in which bosons are initially distributed across both black holes can be analyzed in a similar manner. At $t \to -\infty$, the boson state can be formally expressed as a superposition of molecular orbitals, 
\begin{eqnarray}\label{eq:vector initial condition}
    \ket{\Phi(t=-\infty)}=\ket{\psi^1_{1011}} =\sum_{a=1}^{2N} F_{a}\ket{\sigma^a_v}.
\end{eqnarray}
The coefficient $F_a$ is given by
\begin{eqnarray}\label{eq:vector_initial_condition}
     F_a=\frac{(-1)^{a} \mathbf{D}_a}{\sum_{b=1}^{2N}(-1)^b f_{b0}\mathbf{D}_b},
\end{eqnarray}
where $\mathbf{D}_a=\text{det}(D_{-a})$ and $D_{-a}$ is a matrix obtained by removing the $a$-th row from $D$; and $D$ is defined as in Eq.~\eqref{eq:D_matrix}.

Since the orbit of binary black holes shrinks extremely slowly because of the emission of gravitational waves, the time evolution of the state can be well approximated by an instantaneous Schr\"odinger equation. This implies that the molecular orbitals are time dependent, yet they still form a complete set of bases. Consequently, the boson state at any given time $t$ can be expanded in terms of molecular orbitals, that is, 
\begin{eqnarray}
    \ket{\Phi(t)}=\sum_{a=1}^{2N}F_a(t) e^{-i\int_{-\infty}^t E_a d\tau}\ket{\sigma_v^a}.
\end{eqnarray}
As in the scalar case, the coefficient of each molecular orbital consists of two parts: an exponential factor that describes the free-time evolution, and a term $F_a(t)$ that quantifies the degree of participation of each molecular orbital. The time dependence of these coefficients indicates that bosons may transfer from the primary black hole to the companion. The amount of vector bosons transferred can be evaluated from the coefficients of the atomic orbitals associated with the companion. Those bosons that are transferred to the decaying modes $\ket{\xi^{+1}\psi^2_{n1-1}}$ may be absorbed by the companion black hole, leading to depletion of the boson cloud. Since the atomic orbitals $\ket{\psi^2_{n100}}$, $\ket{\psi^2_{n110}}$ and $\ket{\psi^2_{n120}}$ have nonzero overlap with the molecular orbital $\ket{\sigma_v^a}$, their corresponding coefficients in the state $\ket{\Phi(t)}$ can be found as 
\begin{eqnarray}
    C_{n100}(t) = \frac{\sqrt{6}}{3}C_{n110}(t)=\sqrt{2}C_{n120}(t) = 
    \frac{1}{\sqrt{6}}O(t),
\end{eqnarray}
where
\begin{eqnarray}
    O(t)=\frac{1}{\sqrt{2}}\sum_{a=1}^{2N}F_a(t)f_{a (n+N-1)}(t)e^{-i\int_{-\infty}^t E_a d\tau}. 
\end{eqnarray}
The squared modulus of these coefficients represents the probability of boson transfer to the atomic orbitals $\ket{\psi^2_{n100}}$, $\ket{\psi^2_{n110}}$, and $\ket{\psi^2_{n120}}$, 
\begin{eqnarray}
    |C_{n100}(t)|^2 = \frac{2}{3} |C_{n110}(t)|^2=2 |C_{n120}(t)|^2
    = \frac{1}{6} P(t),
\end{eqnarray}
where 
\begin{eqnarray}
   &&P(t)=\frac{1}{2}\bigg[\sum_{a=1}^{2N}|F_a(t)|^2 f_{a (n+N-1)}(t)^2+\sum_{a\neq b}F_{a}^*(t) F_{b}(t)\nonumber\\
   &&\times f_{a(n+N-1)}(t)f_{b (n+N-1)}(t)\cos\int_{-\infty}^t(E_a-E_b)d\tau \bigg].
\end{eqnarray}
Here, $F_a(t)$ denotes the coefficient corresponding to the $a$-th molecular orbital, with its initial value given by Eq.~\eqref{eq:vector_initial_condition}. The squared modulus $|F_a(t)|^2$ represents the probability that the boson occupies the $a$-th molecular orbital. The lowercase coefficients $f_{a n}$ and $f_{a(n+N-1)}$ denote the contributions from the $n$-th $s$ and $p$ atomic orbitals, respectively. 
The coefficients $f$ can be obtained from Eq.~\eqref{eq: solve_coefficients}, while the energy eigenvalues follow from Eq.~\eqref{eq:determinant}. The time-dependent coefficients $F_a$ can be solved using Eq.~\eqref{eq:non-adiabatic evolution}, in the same manner as in the scalar case.

In the equal-mass case, the $s$-states of the primary and companion black holes combine to form bonding and anti-bonding states, closely analogous to those in the hydrogen molecular ion. These molecular orbitals result from linear combinations of the atomic orbitals $\ket{\psi^1_{1011}}$ and $\ket{\psi^2_{1011}}$, enabling boson transfer as the two black holes approach each other. However, since both of these orbitals correspond to growing modes, the transfer does not contribute to cloud depletion. 
For any other mass ratio, the atomic orbitals $\ket{\psi^1_{1011}}$, $\ket{\psi^2_{n011}}$ and $\ket{\xi^{+1}\psi^2_{np_x}}$ are not degenerate. Consequently, for a given molecular orbital labeled with $a$, the coefficients $f_{a n}$ with $n \ge 1$ are not equal to $f_{a 0}$ at any orbital separation. In the limit of infinite separation, the molecular orbitals reduce to atomic orbitals of either the primary or companion black hole. 
For example, when $a=1$, we find $f_{1 0}=1$ and $f_{1 n}=f_{1(n+N-1)}=0$ for $n \ge 1$, which implies that the molecular orbital $\sigma_v^1$ reduces to the $\ket{\psi^1_{1011}}$ state of the primary black hole. For $a\neq 1$, we have $f_{a 0}=0$, which means that the molecular orbitals $\sigma_v^a$ reduce to atomic orbitals of the companion black hole. According to the initial condition given in Eq.~\eqref{eq:vector_initial_condition}, the only nonzero coefficient is $F_1$ and $F_a=0$ for $a>1$. For $a=1$, we also have $f_{1(n+N-1)=0}$. Therefore, the summations $\sum_{a=1}^{2N}|F_a(t)|^2 f_{a (n+N-1)}(t)^2=0$ and $\sum_{a\neq b}F_a^* F_b$ vanish.
Consequently, $| C_{n100} |^2$, $| C_{n110} |^2$, and $| C_{n120} |^2$ vanish at infinity.

\begin{figure}[htbp]
    \centering
    \includegraphics[width=1\columnwidth]{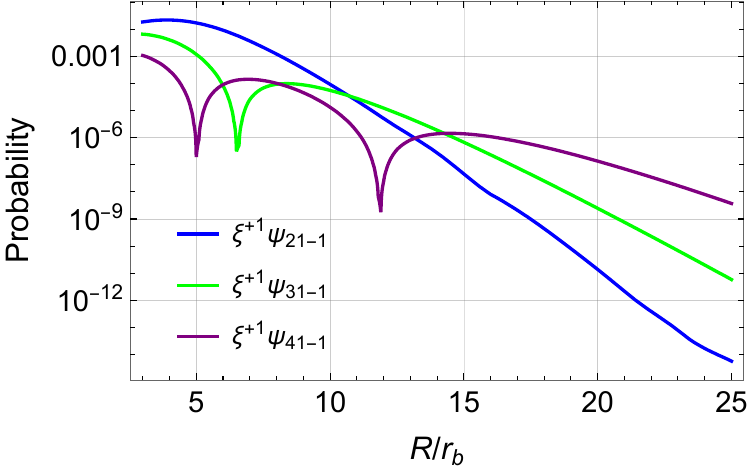}
    \caption{Occupation probabilities of the relevant three states, with $q=2$ and $\alpha=0.1$.
    }
    \label{fig:vector occupation density}
\end{figure}

The gradual shrinkage of the binary black hole orbit modifies the Hamiltonian of the vector bosons, thereby changing their eigenvalues and eigenvectors. As a result, some coefficients that were initially zero, such as $f_{1(n+N-1)}$ and $f_{a1}$ for $a>1$ become nonzero, forming molecular orbitals. Furthermore, bosons may transfer between different molecular orbitals because of non-adiabatic evolution. The initially vanishing coefficient $F_a$ also becomes nonzero. Consequently, the occupation probabilities $|C_{n100}|^2$, $|C_{n110}|^2$ and $|C_{n120}|^2$ gradually increase, showing that vector bosons can transfer to the companion black hole and occupy the decaying modes $\ket{\psi^2_{n100}}$, $\ket{\psi^2_{n110}}$ and $\ket{\psi^2_{n120}}$. 
To illustrate the transfer of bosons to the companion black hole, we evaluate the occupation probabilities of the states $\ket{\xi^{+1}\psi_{21-1}}$, $\ket{\xi^{+1}\psi_{31-1}}$, and $\ket{\xi^{+1}\psi_{41-1}}$ for the case with $q=2$ and $\alpha=0.1$, which are the closest states in energy, as shown in Fig.~\ref{fig:vector occupation density}. In the numerical calculation, we choose a sufficiently large initial orbital separation $R_*=50 r_b$ as a proxy for the infinitely large distance ($t=-\infty$). It can be seen that when the binary black holes are widely separated, no effective molecular orbitals are formed; consequently, all the above occupation probabilities approach zero. As the binary separation decreases, the atomic orbitals begin to overlap, leading to a gradual increase in the corresponding occupation probabilities. The trend is qualitatively similar for other values of $q$ and $\alpha$. 

Vector bosons occupying the decaying modes of the companion black hole could be absorbed by it, resulting in the depletion of the vector cloud. In our analysis, we neglect both the back reaction of boson transfer on the binary's orbital evolution and its influence on the decay rate. The decay process involves three decaying modes, labeled as $\ket{\psi^2_{n1j0}}$ with $j \in \{0,1,2\}$, and their occupation probabilities are $|C_{n1j0}|^2$. The overall depletion rate of the vector cloud is therefore given by
\begin{eqnarray}\label{eq:total depletion}
\frac{dM_c}{dt}=2\sum_{n=2}^N\sum_{j=0}^2|C_{n1j0}|^2\Gamma_{n1j0}M_c.
\end{eqnarray}

\begin{figure}[htbp]
    \centering
    \includegraphics[width=1\columnwidth]{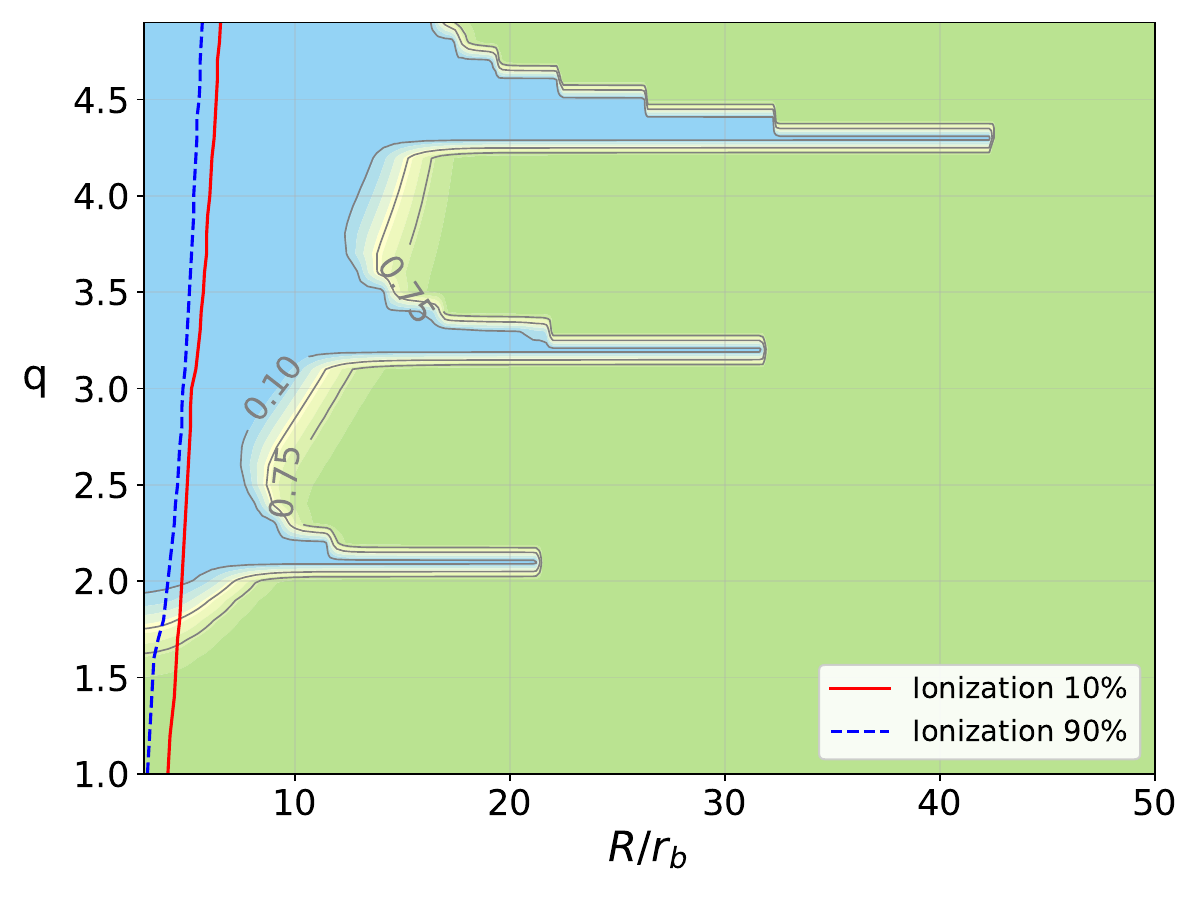}
    \caption{Vector cloud depletion by molecular orbits and the dominant separation of ionization. The green region indicates no cloud depletion, the blue region represents sharp cloud depletion, and the red curve corresponds to ionization removing approximately $10\%$ of the cloud, another blue curve represent $90\%$ ionization of the cloud. We choose the fine-structure-like constant $\alpha=0.1$.
    }
    \label{fig:vector total depletion}
\end{figure}

We investigate vector boson cloud depletion in binary black hole systems with varying mass ratios and evaluate the fraction $M_c/M_0$, as shown in Fig.~\ref{fig:vector total depletion} for $\alpha=0.1$. We have neglected other dynamical effects, such as self-annihilation and Bohr mixing, and focused on the depletion caused only by boson transfer. We further assume that the boson cloud exists at orbital separations $R_* \geq 50 \, r_b$. The contribution of decaying modes with different principal quantum numbers $n$ to cloud depletion depends on the mass ratio. We account for the quantum numbers from $n=1$ to $5$, which are the most relevant and dominate the transfer process for the mass ratio $q$ in the regime between $1$ and $5$. 

The vector cloud, similar to its scalar counterpart, can be completely absorbed by the companion black hole when the mass ratio is sufficiently large. Moreover, there exist special mass ratios in which the depletion extends to larger separations compared to other cases. Such special mass ratios typically correspond to integer values. At infinite separation of the two black holes, the energy of the atomic orbital with principal quantum number $n$ of the companion black hole, relative to that of the primary’s $\ket{\psi^1_{1011}}$ orbital, is given by $q/n$. Consequently, when the mass ratio $q$ matches $n$, molecular orbitals can form at larger separations. When the mass ratio is very small, the energy eigenvalue of the state $\ket{\psi^2_{21j0}}$ in the companion differs significantly from that of the primary state $\ket{\psi^1_{1011}}$. Therefore, molecular orbitals cannot be formed effectively, and the cloud cannot be completely depleted into the companion. Since the dominant growing mode $\ket{\psi^1_{1011}}$ of the vector boson in the primary black hole is more tightly bound than that of the scalar boson, and the lowest principal quantum number of the decaying levels in the companion black hole is $n=2$, the mass ratio required for vector bosons to be completely depleted by the companion black hole must be larger than that for scalar bosons.

\subsection{Cloud depletion due to other dynamical effects}

For the vector boson, the dominant growing mode $\ket{\psi^1_{1011}}$ has a higher binding energy than its scalar counterpart. As a result, the ionization occurs at smaller orbital separations. We calculate the typical orbital separations at which cloud depletion due to ionization occurs, as shown in Fig.~\ref{fig:vector total depletion}. In the numerical calculation, we assume that the cloud exists when the separation is approximately $R_*=50 \, r_b$. For sufficiently large mass ratios, the typical orbital separations at which cloud depletion occurs through mass transfer are larger than those for ionization, implying that mass transfer dominates the cloud depletion. However, for small mass ratios, the formation of molecular orbitals is inefficient, and only a tiny fraction of the boson cloud can be depleted through mass transfer. In this regime, the effect of ionization becomes dominant. We further vary the parameter $\alpha$ and find a similar trend: mass transfer dominates at large mass ratios, whereas ionization dominates at small mass ratios.

Gravitational perturbation from the companion cannot induce hyperfine mixing when vector bosons occupy the state $\ket{\psi^1_{1011}}$ of the primary black hole. For co-rotating orbits, neither hyperfine nor Bohr resonance is present. The low resonance rates combined with weak decay rates result in a tiny amount of cloud depletion into the primary, so the mass transfer to the companion dominates the cloud depletion. For counter-rotating orbits, only Bohr resonance occurs. However, the decay rates of relevant decaying modes remain very low, leading to a small amount of cloud depletion into the primary black hole. Therefore, most of the vector boson cloud is either depleted into the companion through mass transfer or ionized, depending on the mass ratio.

\section{Arbitrary spin orientations}\label{Sec: Any_orientations}

In the above discussions, the spins of two black holes are assumed to be aligned (parallel). However, there is no preferred direction for the black hole spin. In this section, we propose a method to construct molecular orbitals in a binary black hole system, in which the relative spin orientation between the two black holes is arbitrary. It is found that the relative spin orientation has an important effect on the cloud depletion. 
\begin{figure}[htbp]
    \includegraphics[width=1\columnwidth]{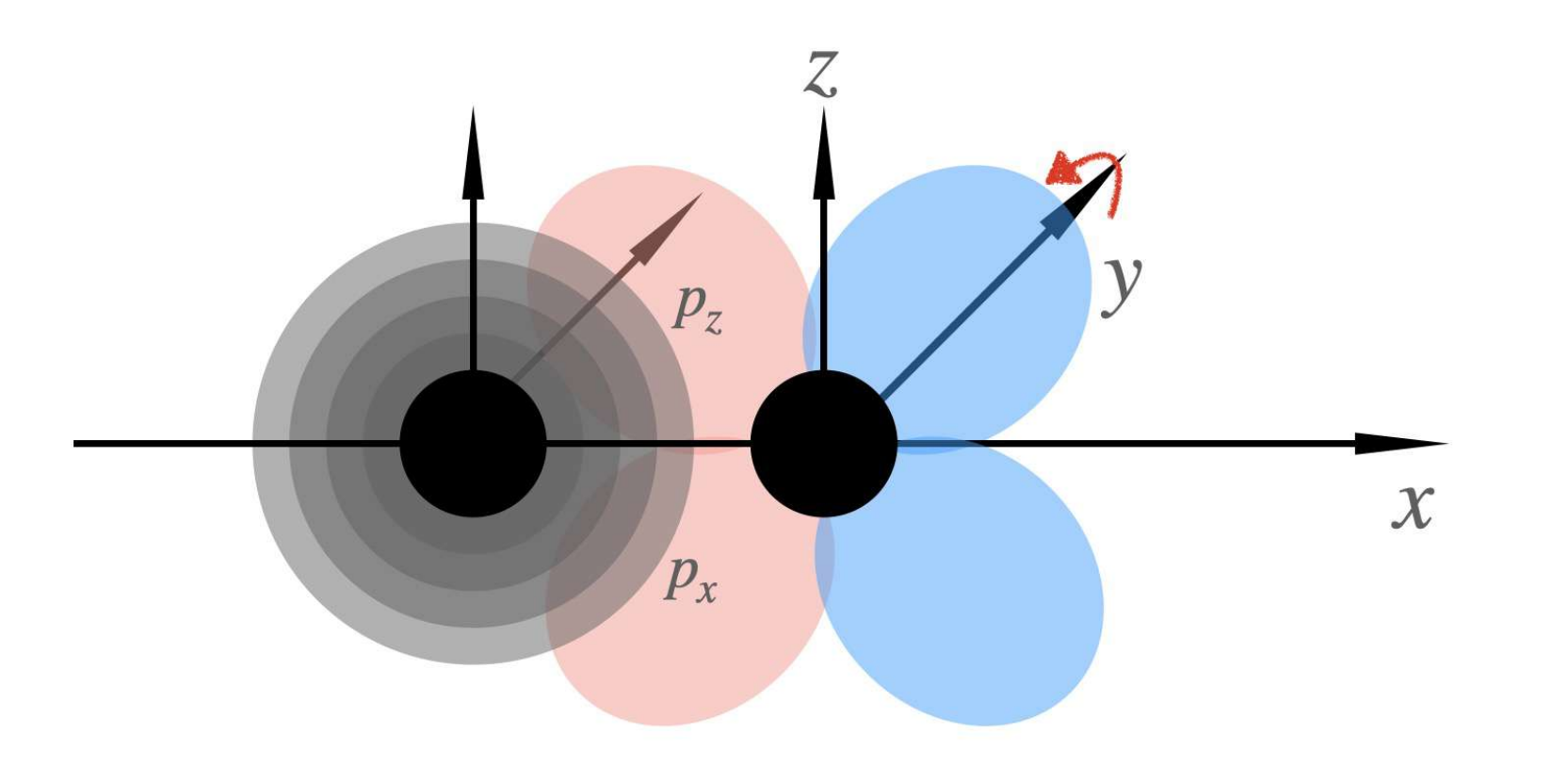}
    \caption{Illustration of the wavefunction overlap after rotation. The primary black hole is surrounded by a vector cloud occupying the $\ket{1011}$ state. The companion black hole hosts wavefunctions in the $p_x$ and $p_z$ states, which overlap with the primary's cloud. The degree of overlap varies with the relative orientation of the companion, influencing the efficiency of cloud depletion.}
    \label{fig: rotation wavefunc overlap}
\end{figure}

\subsection{Molecular orbitals for arbitrary spin orientation}

We neglect the effects of spin-orbit and spin-spin interactions between two black holes. The role of spin is only to define a preferred $z$ direction for the orbital wave functions. When the companion black hole approaches the primary black hole, bosons inevitably transfer to the companion black hole due to gravitational perturbations. In the case of parallel spin, the atomic orbitals of the primary black hole and certain atomic orbitals from the companion can be combined to form molecular orbitals. Suppose that the spin of the companion black hole is rotated, it is then expected that the atomic orbital of the primary black hole can be combined with a specific linear combination of atomic orbitals from the companion form molecular orbitals. The specific linear combination is determined by a three-dimensional rotation transformation. Since the decay rate for different orbitals is different, it is expected that the cloud depletion rate would be different for different spin orientations of the companion black hole.

When the spin-orbit and spin-spin interactions are neglected, the Hamiltonian of the boson is independent of the relative spin orientation between two black holes. This implies the existence of molecular orbitals that preserve the spatial symmetries encoded in Eq.~\eqref{eq:Hamiltonian}. Therefore, we can similarly construct the $\sigma$ and $\pi$ orbitals for the scalar bosons as
\begin{eqnarray}
\ket{\sigma^a_s}&=&f_{a0}\ket{\psi^1_{2p_x}}+\sum_{n=1}^{N}f_{an}\ket{\psi^2_{ns}(\mathbf{r})} \nonumber\\ 
&&
+\sum_{n=2}^N f_{a(N+n-1)}\ket{\psi^2_{np_x}(\mathbf{r})},\nonumber\\
\ket{\pi^\mu_s}&=&g_{\mu1}\ket{\psi^1_{2p_y}}+\sum_{n=2}^N g_{\mu n}\ket{\psi^2_{np_y}(\mathbf{r})},
\end{eqnarray}
and $\sigma$ molecular orbitals for vector bosons as 
\begin{eqnarray}
\ket{\sigma^i_v}&=&c_{i0}\ket{\psi^1_{1011}}+\sum_{n=1}^N c_{i n}\ket{\psi^2_{n011}(\mathbf{r})}\nonumber\\
&&+\sum_{n=2}^N c_{i(n+N-1)}\ket{\xi^{+1}\psi^2_{np_x}(\mathbf{r})}.
\end{eqnarray}
Here, the atomic wave functions of the companion black hole are expressed in a reference coordinate system $\mathbf{r}=(r,\theta,\phi)$, in which the $y$- and $z$-axes are parallel to those of the primary black hole, and the $x$-axis is aligned with the primary’s $x$-axis. The reference coordinate system is established only for convenience, as will become clear in the following. We also introduce a coordinate system $\mathbf{r'}=(r',\theta',\phi')$ that co-rotates with the companion's spin orientation, that is, the $z'$-axis is aligned with the spin orientation. These two coordinate systems are related by a 3-dimensional spatial rotation. In particular, the vector $\mathbf{r'}$ is related to $\mathbf{r}$ via,
\begin{eqnarray}
\mathbf{r'}=\mathcal{R}(\zeta,\beta,\gamma) \, \mathbf{r},
\end{eqnarray}
where $\mathcal{R}(\zeta,\beta,\gamma)$ is the rotation matrix specified by the Euler angles $(\zeta, \beta, \gamma)$.

Our objective is to express the atomic orbitals in the reference coordinate system in terms of linear combinations of atomic orbitals in the corotated coordinate system. It is evident that the $s$ orbitals remain unchanged under spatial rotation because of their spherical symmetry. However, $p$ orbitals generally change under spatial rotation. To determine the transformation of atomic orbitals, we need to consider transformations of spherical harmonics under the spatial rotation, which is given by 
\begin{eqnarray}\label{eq:HarmTransform}
    Y_{1m}(\mathbf{r'}) &=& \sum_{n=-1}^1 D_{mn}^1 (\zeta,\beta,\gamma)Y_{1n}(\mathbf{r})\nonumber\\
    &=&
    \sum_{n=-1}^1 \big[ D^1_{x}(\zeta)D^1_{y}(\beta)D^1_{z}(\gamma) \big]_{mn} Y_{1n}(\mathbf{r})\nonumber\\
    &=& 
    \sum_{n, k,p =-1}^1 
    D_{x,mk}^1(\zeta) D_{y,kp}^1(\beta) 
    D^1_{z,pn}(\gamma) Y_{1n}(\mathbf{r}), \nonumber\\
\end{eqnarray}
where $D_{mn}^1(\zeta,\beta,\gamma)$ denote the matrix elements of the Wigner-$D$ matrix, with $m$ and $n$ representing the magnetic quantum numbers. It is well known that an arbitrary 3-dimensional spatial rotation can be decomposed into: (1) a rotation about the $z$-axis by an angle $\gamma$ and then (2) a rotation about the initial $y$-axis by an angle $\beta$, and finally (3) a rotation about the new $x$-axis by an angle $\zeta$. These three spatial rotations with respective to specific coordinate axes give rise to a decomposition of the Wigner-$D$ operator as $D^1(\zeta,\beta,\gamma) =  D^1_x(\zeta) D^1_y(\beta) D^1_z(\gamma)$, where the subscripts $x, y$ and $z$ indicate the corresponding axes. 
The rotation $D^1_z(\gamma)$ only induces a phase or a trivial unitary transformation within the $(p_x, p_y)$ subspace, which does not affect the overlap amplitudes or other rotationally invariant quantities considered here. Therefore, the dependence on $\gamma$ can be safely omitted.
For clarity, we emphasize that the Euler angles $(\zeta, \beta, \gamma)$ used here differ from the conventional $z$-$y$-$z$ Euler parametrization $(\gamma, \beta, \gamma)$. Our notation is chosen deliberately to reflect the actual physical rotation sequence: a rotation about the new $x$-axis at the final step rather than a second $z$-axis rotation. 

In order to express the wave functions defined in the reference coordinate system in terms of the rotated basis functions that co-move with the companion black hole's spin direction, we invert the transformation in Eq.~\eqref{eq:HarmTransform} and obtain 
\begin{eqnarray}
    Y_{1n}(\mathbf{r})=\sum_{m=-1}^1\Big[ D^1 (\zeta,\beta,\gamma)^{-1}\Big]_{n m}Y_{1m}(\mathbf{r'}).
\end{eqnarray} 

When considering only rotations about the $y$- and $x$-axes, the spherical harmonics $Y_{1n}$ in the reference frame $\mathbf{r}$ can be expressed as
\begin{eqnarray}\label{eq: Spherical_rot}
    Y_{1n}(\mathbf{r})=\sum_{m,k=-1}^1 [D^{1}_{y}(\beta)^{-1}]_{nk}[D^{1}_{x}(\zeta)^{-1}]_{km}Y_{1m}(\mathbf{r'}), 
\end{eqnarray}
where the indices $\{1,0,-1\}$ label the first, second, and third rows (or columns) of the matrices. 
The corresponding Wigner-$D$ matrices are given by
\begin{eqnarray}
\label{eq: Dx}
D^1_x(\zeta)=\frac{1}{2}\left(\begin{matrix}
   1+\cos\zeta & -i\sqrt{2}\sin\zeta & \cos\zeta-1 \\
   -i\sqrt{2}\sin\zeta& 2\cos\zeta & -i\sqrt{2}\sin\zeta\\
   \cos\zeta-1 & -i\sqrt{2}\sin\zeta & 1+\cos\zeta
\end{matrix}\right),
\end{eqnarray}
and
\begin{eqnarray}
    D^1_y(\beta)=\frac{1}{2} \begin{pmatrix}
   1+\cos\beta & -\sqrt{2} \sin \beta & 1-\cos\beta \\
   \sqrt{2} \sin \beta & 2 \cos \beta & -\sqrt{2} \sin\beta \\
   1-\cos\beta & \sqrt{2} \sin \beta & 1+\cos\beta
\end{pmatrix}.
\end{eqnarray}
The companion's wave function $\psi^2_{np_x}(\mathbf{r})$ in the reference frame can be expressed as
\begin{eqnarray}\label{eq: rotation-p-state}
    \psi^2_{np_x}(\mathbf{r}) &=& \frac{1}{\sqrt{2}}[\psi^2_{n1-1}(\mathbf{r})-\psi^2_{n11}(\mathbf{r})]\nonumber\\
    &=&\frac{1}{\sqrt{2}}\sum_{m=-1}^1\bigg\{\sum_{k=-1}^1 [D^1_{y}(\beta)^{-1} ]_{(-1)k} [D^1_{x}(\zeta)^{-1}]_{km}\nonumber\\
    &&-\sum_{k=-1}^1[D^1_{y}(\beta)^{-1}]_{1k}[D^1_{x}(\zeta)^{-1}]_{km}\bigg\}\psi^2_{n1m}(\mathbf{r'})\nonumber\\
    &=&\cos\beta \, \psi^2_{np_x}(\mathbf{r'})+\sin\beta\sin\zeta \, \psi^2_{np_y}(\mathbf{r'})\nonumber\\
    &&-\cos\zeta\sin\beta \, \psi^2_{np_z}(\mathbf{r'}).
\end{eqnarray}
This explicitly shows how the atomic states $\psi^2_{np_x}(\mathbf{r'})$, $\psi^2_{np_y}(\mathbf{r'})$, and $\psi^2_{np_z}(\mathbf{r'})$ in the co-rotated coordinate system are linearly combined to form the state $\psi^2_{np_x}(\mathbf{r})$ in the reference coordinate system after two successive rotations: first by an angle $\beta$ about the $y$-axis, and then by an angle $\zeta$ about the new $x$-axis.

We can now compute the coefficients of the molecular orbitals after rotating the companion’s spin orientation. 
However, since the rotation of the companion black hole's spin axis does not alter the Hamiltonian, and the overlap between the newly constructed wave function $\psi^2_{np_x}(\mathbf{r})$ and the primary black hole's wave function remains unchanged (see Appendix~\ref{Appendix:spin_orientation}), the coefficients $f_{an}$ are identical to those in the parallel-spin case. The only difference is that the coefficient preceding the rotated wave function $\psi(\mathbf{r'})$ needs to be multiplied by a rotation-angle-dependent factor, as indicated in Eq.~\eqref{eq: rotation-p-state}

As a concrete example, we consider vector bosons whose decay mode corresponds exclusively to the $np_x$ orbital. Unlike the scalar case, where the $s$-state dominates and is insensitive to spin orientation, the overlap of the $np_x$ orbital with the atomic orbitals of the primary black hole and thus the resulting cloud depletion depends sensitively on the spin direction.

\subsection{Depletion of vector boson cloud in any orientation}

We now investigate the dynamics of the vector boson cloud in a binary black hole system, where the companion’s spin is rotated by Euler angles $\zeta$ and $\beta$. We again assume that the vector bosons initially occupy only the $\ket{\psi^1_{1011}}$ state of the primary black hole, while the companion contains no bosons.  At $t \to -\infty$, the boson state can be formally expressed as a superposition of molecular orbitals, 
\begin{eqnarray}
    \ket{\Phi(t=-\infty)}=\ket{\psi^1_{1011}} =\sum_{a=1}^{2N} F_{a}\ket{\sigma^a_v}.
\end{eqnarray}
The coefficient $F_a$ is also given by Eq.~\eqref{eq:vector_initial_condition} and is independent of the rotation angle. 
Similarly, the boson state at any given time $t$ can be expanded in terms of molecular orbitals at time $t$, that is,
\begin{eqnarray}
    \ket{\Phi(t)}=\sum_{a=1}^{2N}F_a(t) e^{-i\int_{-\infty}^t E_a d\tau}\ket{\sigma_v^a(t)}.
\end{eqnarray}

The state $\ket{\xi^{+1}\psi_{np_x}}$ is made up of decaying modes $\ket{\psi_{n100}}$, $\ket{\psi_{n110}}$, $\ket{\psi_{n120}}$, whose combination coefficients depend on rotation angles, as described by Eq.~\eqref{eq: rotation-p-state}. Therefore, the coefficient of each decaying mode is also a function of rotation angles given by
\begin{eqnarray}
    C_{n100}(t, \zeta, \beta)&=&\frac{\sqrt{6}}{3}C_{n110}(t, \zeta, \beta)=\sqrt{2} \, C_{n120}(t, \zeta, \beta)\nonumber\\
    &=&\frac{1}{\sqrt{6}}(\cos\beta+i \sin\beta\sin\zeta)O(t),
\end{eqnarray}
where
\begin{eqnarray}
    O(t)=\frac{1}{\sqrt{2}}\sum_{a=1}^{2N}F_a(t)f_{a (n+N-1)}(t)e^{-i\int_{-\infty}^t E_a d\tau}.
\end{eqnarray}
The squared modulus of these coefficients represents the probability of boson transfer to the atomic orbitals $\ket{\psi^2_{n100}}$, $\ket{\psi^2_{n110}}$, and $\ket{\psi^2_{n120}}$, 
\begin{eqnarray}
    |C_{n100}(t,\zeta,\beta)|^2 
    &=& \frac{2}{3} |C_{n110}(t,\zeta,\beta)|^2=2 |C_{n120}(t,\zeta,\beta)|^2 \nonumber\\
    &=&\frac{1}{6}\big(\cos^2\beta +\sin^2\beta \sin^2 \zeta \big) P(t),
\end{eqnarray}
where 
\begin{eqnarray}
   P(t) &=& \frac{1}{2}\bigg[\sum_{a=1}^{2N}|F_a(t)|^2 f_{a (n+N-1)}(t)^2\nonumber\\
   &&+\sum_{a\neq b}F_{a}^*(t) F_{b}(t) f_{a(n+N-1)}(t)f_{b (n+N-1)}(t)\nonumber\\
   &&\times\cos\int_{-\infty}^t (E_a-E_b) d\tau\bigg]. 
\end{eqnarray}
Although the molecular-orbital coefficients $F$ and the atomic orbital coefficients $f$ are independent of rotation angles, the total occupation probability remains scaled by an additional factor $\cos^2\beta+\sin^2 \beta \sin^2 \zeta$. When $\zeta=0, \pi, 2\pi$ and $\beta=\pi/2, 3\pi/2$, the probability vanishes. In this case, $\ket{\xi^{+1}\psi_{np_x}(\mathbf{r})}$ is entirely composed of $\ket{\xi^{+1}\psi_{np_z}(\mathbf{r'})}$, which is a growing mode, and thus the vector cloud does not decay. In contrast, when $\beta=0, \pi, 2\pi$, the factor $\cos^2 \beta +\sin^2 \beta \sin^2 \zeta$ reaches its maximum, and consequently the occupation probability reaches its maximum.

In our analysis, we neglect the back-reaction of boson transfer on the binary’s orbital evolution, as well as its impact on the decay rate. Vector bosons deplete through three decaying modes, denoted as $\ket{\psi^2_{n1j0}}$ with $j \in {0,1,2}$, the occupation probabilities of which are $|C_{n1j0}|^2$. Consequently, the total depletion rate of the vector cloud can be written as
\begin{eqnarray}\label{eq:total depletion}
\frac{dM_c}{dt}=2\sum_{n=2}^N\sum_{j=0}^2|C_{n1j0}(\zeta,\beta)|^2\Gamma_{n1j0}M_c.
\end{eqnarray}
In order to investigate the effects of the relative spin orientation between the two black holes on the boson cloud depletion, we compute $M_c/M_0$ as a function of the orbital separation and rotation angles.

\begin{figure}
    \centering
    \includegraphics[width=0.9 \linewidth]{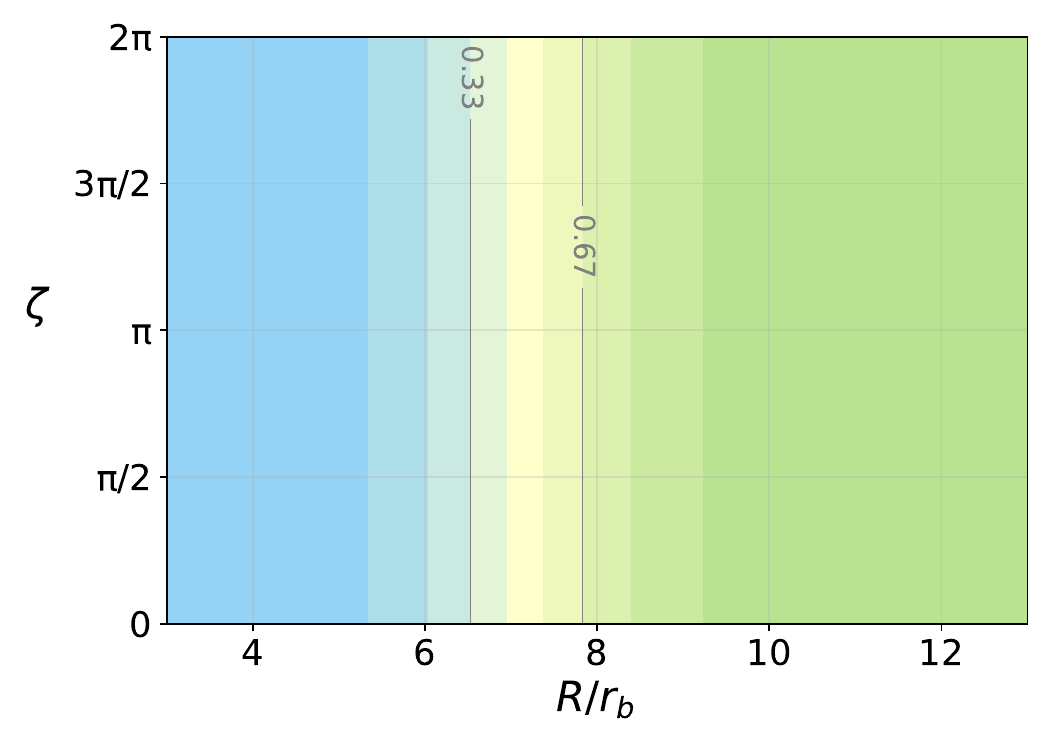}
    \caption{Boson cloud depletion with the companion black hole only rotating about $x$-axis. The green region indicates no cloud depletion and the blue region represents sharp cloud depletion. We choose $q=2$ and $\alpha=0.1$.}
    \label{figure: vector_rot_x}
\end{figure}

\begin{figure}
    \centering
    \includegraphics[width=0.9 \linewidth]{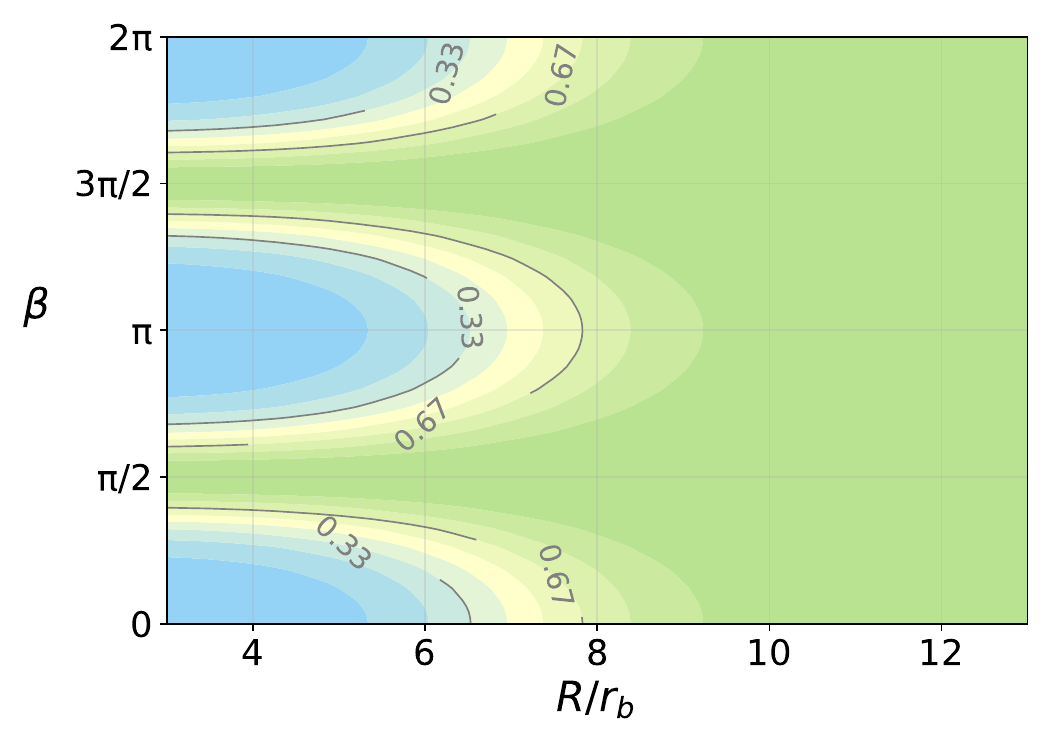}
    \caption{Boson cloud depletion with the companion black hole only rotating about $y$-axis. The green region indicates no cloud depletion and the blue region represents sharp cloud depletion. We choose $q=2$ and $\alpha=0.1$. }
    \label{figure: vector_rot_y}
\end{figure}

When the companion black hole rotates solely about the $x$-axis, namely, $\beta=0$, it is evident that the probabilities of occupying the decaying modes $\ket{\psi^2_{n1j0}}$ remain the same as when the spin of the companion black hole is aligned with the primary. Figure~\ref{figure: vector_rot_x} shows $M_c/M_0$ as a function of orbital separation and angle $\zeta$ given that 
$\beta=0$, with $q=2$ and $\alpha = 0.1$. This is because, in this configuration, the corresponding atomic state of the companion is identical to that in the spin-parallel case, $\ket{\xi^{+1}\psi_{np_x}(\mathbf{r'})}=\ket{\xi^{+1}\psi_{np_x}(\mathbf{r})}$, implying that the wave functions of the companion black hole involved in the formation of molecular orbitals remain unchanged.
When the companion black hole rotates solely about the $y$-axis, namely $\zeta=0$, the probabilities of occupying the decaying modes $\ket{\psi^2_{n1j0}}$ vary as $\cos^2 \beta$, indicating angle-dependent cloud depletion. Figure~\ref{figure: vector_rot_y} shows $M_c/M_0$ as a function of orbital separation and angle $\beta$ given that 
$\zeta=0$, with $q=2$ and $\alpha = 0.1$. When $\beta$ is within the range $(0, \pi/2)$, the state $\ket{\xi^{+1}\psi^2_{np_x}(\mathbf{r})}$ gains a large contribution from $\ket{\psi^2_{np_z}(\mathbf{r'})}$ while the contribution from $\ket{\psi^2_{np_x}(\mathbf{r'})}$ decreases. This reduces the participation of the decaying modes and, consequently, diminishes the depletion of the cloud. In the range $(\pi/2, \pi )$, the situation is reversed—the contribution of $\ket{\psi^2_{np_x}}$ increases, causing the cloud to deplete efficiently.

\section{Effect on gravitational wave radiation}
\label{Sec: Gravitational_waves}

Transfer of bosons from the primary black hole to the companion alters the mass distribution within the binary black hole system, thereby affecting its gravitational wave signal. Gravitational-wave modifications induced by the presence of bosons, such as resonance mixing, have already been extensively studied~\cite{Peng:2025zca, Miller:2025yyx, Aggarwal:2020olq}. In this section, we consider the resulting offset in the gravitational wave power of the binary. The average power of gravitational radiation can be calculated via
\begin{eqnarray}
    \frac{dE}{dt}=\frac{1}{5} \sum_{ij} \left<\dddot{Q}_{ij} \dddot{Q}^{ij}\right>,
\end{eqnarray}
where $Q_{ij}$ is the quadrupole moment
\begin{eqnarray}
    Q_{ij}=\int dx^3 \rho(\mathbf{r})\Big(x_i x_j-\frac{1}{3}x^2 \delta_{ij}\Big),
\end{eqnarray}
with $\rho(\mathbf{r})$ the mass density distribution. Consider a binary system where the primary black hole has mass $M_1$ and the companion has mass $M_2 = q M_1$. When the orbital separation is much larger than their gravitational radii, the binary can be well approximated as two point masses. The average radiation power is given by
\begin{eqnarray}
P_0=\frac{32}{5}\bigg(\frac{M_1 M_2}{M_1+M_2}\bigg)^2 R_*^4 \, \Omega^6,
\end{eqnarray}
where $R_*$ is the orbital separation and $\Omega$ is the orbital frequency. 
When the primary black hole develops a boson cloud, bosons are dispersed around it with density $\rho(\mathbf{r})$ and a characteristic length scale $r_b$. In this case, the boson cloud and the primary black hole can no longer be approximated as a single point mass. After the molecular orbitals are formed, the boson cloud may transfer and occupy the atomic states of the companion black hole, thereby further influencing the mass distribution. 

To illustrate the offset in radiation power of gravitational waves between a binary with a boson cloud and one without, we consider the following setup. In the case where the system contains a boson cloud, the primary black hole has mass $M_1$, initially accompanied by a boson cloud of mass $\alpha M_1$, and the companion black hole has mass $M_2$. In subsequent evolution, the boson cloud is redistributed within the binary system and could be absorbed by both the primary and the companion black holes, resulting in time-dependent mass distribution. In the case where the system doses not contain a boson cloud, we assume that the primary black hole instead has mass $M_1+\alpha M_1$ and the companion black hole again has mass $M_2 = \alpha M_1$.
We further assume that the system with a boson cloud has the same orbital frequency as the corresponding binary without cloud. When the orbital separation of the companion black hole becomes comparable to the characteristic scale of the boson cloud, mass transfer from the cloud can drive a rapid inspiral of the companion black hole~\cite{Guo:2025ckp,Guo:2024iye}.

\begin{figure}[htbp]
    \centering
    \includegraphics[width=1\columnwidth]{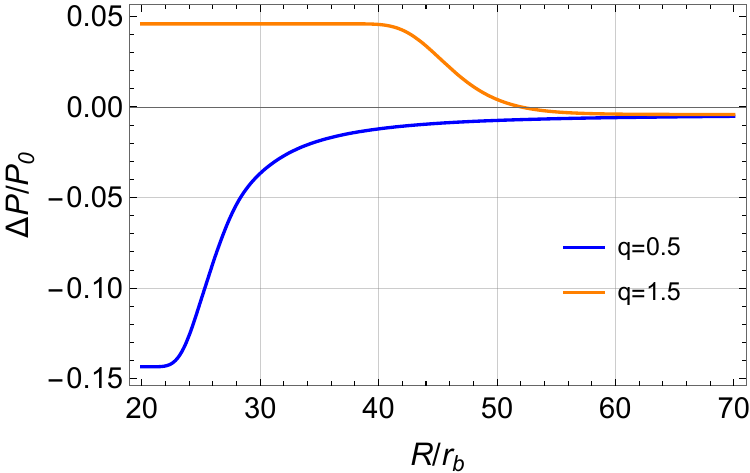}
    \caption{Power offset in gravitational waves due to scalar cloud mass transfer, with $\alpha=0.1$. The orange region represents the case where $q=1.5$. The blue region indicates the deviation in gravitational wave power where $q=0.5$.
    }
    \label{fig:scalar gws}
\end{figure}

When the bosons are transferred to and absorbed by the companion, the mass distribution of the system is altered, resulting in a modification of the quadrupole moment and, consequently, of the emitted gravitational waves. We denote by $\Delta P =P_1-P_0$ the change in the average power of gravitational radiation induced by mass transfer and cloud depletion, where $P_1$ is the average power with a boson cloud, $P_0$ is the power without a boson cloud.

Figure~\ref{fig:scalar gws} shows the fractional change in gravitational-wave power, $\Delta P / P_0$, induced by scalar boson transfer. When the companion black hole is far from the primary, mass transfer has not yet occurred. In this regime, the small deviation in the gravitational-wave power arises from the mass distribution of the boson cloud surrounding the primary black hole, with $\rho \propto |\psi^1_{211}|^2$. As the companion black hole approaches the primary, the boson cloud begins to transfer to and be absorbed by the companion, leading to an increase in its mass. Meanwhile, the primary black hole also gains a fraction of the boson cloud through level mixing. At this stage, the occupation probability of the decaying levels around the companion black hole remains low, so the change in gravitational-wave power caused by the boson cloud distributed outside the companion black hole is not significant. As the transfer of the boson cloud continues, the mass of the companion black hole steadily increases, while the mass of the primary black hole also grows, but to a lesser extent. Eventually, the boson cloud is fully absorbed, and the difference of the gravitational-wave power asymptotically approaches a stable value.

\begin{figure}[htbp]
    \centering
    \includegraphics[width=1\columnwidth]{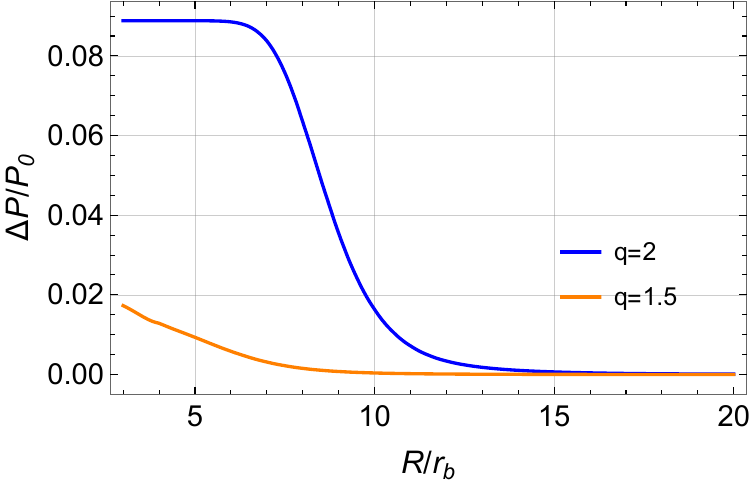}
    \caption{Power offset in gravitational waves due to vector cloud mass transfer, with $\alpha=0.1$. The orange region represents the regime where $q=1.5$ . The blue region indicates the deviation $q=2$.
    }
    \label{fig:vector gws}
\end{figure}

In Fig.~\ref{fig:vector gws}, we present the fractional change in gravitational-wave power induced by the transfer of vector bosons. Since we consider the $\ket{1011}$ energy level, the spatial distribution of which is spherically symmetric, no power difference appears at large orbital separations. Compared with scalar bosons, mass transfer of the vector cloud occurs only when the companion black hole is at a much closer distance, and consequently the deviation in the gravitational-wave amplitude also emerges at smaller orbital separations. We show results for two mass ratios, $q=1.5$ and $2$. For a mass ratio of $q=2$, the boson cloud initially leads to increases in both the companion black hole mass and the gravitational-wave power shift; it is then completely absorbed, after which the power shift reaches a steady value. In contrast, for a mass ratio of $q=1.5$, although the companion black hole can absorb the vector boson cloud, it cannot fully accrete the cloud before ionization occurs; therefore, only the rising phase of the power shift is shown in this figure.

\section{Conclusions}
\label{Sec: Conclusion}
We have investigated the mass transfer of both scalar and vector boson clouds in a binary black hole system, and the subsequent cloud depletion into the companion black hole, with particular attention to mass ratios close to unity and arbitrary spin orientations of the companion black hole. By assuming that all bosons initially occupy the fastest-growing mode of the primary, we find that mass transfer from the scalar cloud typically occurs at larger orbital separations compared to the vector case. As the mass ratio varies, the atomic state of the companion black hole that dominates the mass transfer also changes.

For sufficiently larger mass ratios, both the scalar and vector bosons are transferred to the companion and subsequently absorbed by it, leading to a complete depletion of the boson cloud before the ionization occurs. For certain mass ratios where the energy levels of the primary and companion become almost degenerate, molecular orbitals can be formed more effectively, resulting in cloud depletion at larger orbital separations. However, when the companion black hole is sufficiently lighter than the primary, the molecular orbitals fail to form effectively, and the bosons cannot transfer and deplete before it is ionized.

We compare the mass depletion resulting from mass transfer with that arising from hyperfine and Bohr mixings. For the scalar cloud, regardless of whether the binary is co-rotating or counter-rotating, the dominant depletion channel is the hyperfine mixing into the primary black hole, with only a small fraction of the cloud transferring to the companion. In contrast, the vector cloud does not exhibit hyperfine resonance and the Bohr mixing leads only to a minor depletion. As a result, the vector cloud predominantly transfers into the companion black hole in most circumstances. When the mass ratio becomes sufficiently small, mass transfer ceases for both scalar and vector clouds, and ionization becomes the dominant depletion mechanism as the binary separation decreases.
We further classify the effects of the companion black hole's spin orientation on the cloud depletion. For the scalar cloud, when the dominant decaying mode is the $s$-state, the depletion is unaffected by the spin orientation because of the spherical symmetry of the $s$-orbitals. In contrast, for the vector cloud, the cloud depletes at larger separations in the spin-parallel configuration compared to other spin orientations.

The framework presented in this work focuses on scenarios where the mass ratio is close to unity. As the mass ratio increases by several orders of magnitude, the Born-Oppenheimer approximation may no longer be valid, and alternative methods need to be developed to accurately describe the dynamics of the boson cloud. The mass transfer of both scalar and vector clouds leads to a redistribution of mass and angular momentum within the binary system, potentially influencing orbital evolution. A more detailed investigation of these effects will be addressed in future work. 

{\bf Acknowledgements:} D. S. would like to acknowledge the support from the Fundamental Research Funds for the Central Universities, HUST (Grant No. 5003012068), Wuhan Young Talent Research Funds (Grant No. 0106012013), and the National Key R\&D Program of China ``Gravitational Wave Detection" (Grant No. 2023YFC2205800).

\appendix

\section{Wave function}

For a massive scalar field $\Phi$ with the Lagrangian
\begin{eqnarray}
    \mathcal{L}=\frac{1}{2}(\partial_a \Phi) (\partial^a \Phi) -\frac{1}{2}\mu^2 \Phi^2,
\end{eqnarray}
the equation of motion takes the form
\begin{eqnarray}
    (g^{ab}\nabla_a\nabla_b-\mu^2)\Phi=0,
\end{eqnarray}
where $g^{ab}$ is the metric of the background spacetime. In the background Kerr metric, when the Compton wavelength of the field is larger than the size of the black hole, the system admits non-relativistic, hydrogen-like bound state solutions. The scalar field can be expressed as
\begin{eqnarray}
    \Phi(t,\mathbf{r})=\frac{1}{\sqrt{2\mu}}\big[\psi(t,\mathbf{r})e^{-i\mu t}+\psi^*(t,\mathbf{r})e^{i\mu t} \big],
\end{eqnarray}
where $\psi(t,\mathbf{r})$ is a complex scalar field that approximately satisfies the Schr\"{o}dinger equation with a Coulomb-like potential,
\begin{eqnarray}
    i\frac{\partial}{\partial t}\psi(t,\mathbf{r})=\bigg(-\frac{\nabla^2}{2\mu}-\frac{\alpha}{r} \bigg) \psi(t,\mathbf{r}).
\end{eqnarray}
The eigenstates $\psi_{nlm}$ are approximately given by hydrogenic wave functions,
\begin{eqnarray}
    \psi_{nlm}(t,\mathbf{r})\simeq e^{-i(\omega-\mu)t}R_{nl}(r)Y_{lm}(\theta,\phi),
\end{eqnarray}
where $n$ is the principal quantum number, $l$ is the angular momentum quantum number, and $m$ is the azimuthal quantum number.

For a free massive spin-1 vector field, the corresponding Lagrangian is
\begin{eqnarray}
\mathcal{L_A}=-\frac{1}{4}F_{\mu\nu}F^{\mu\nu}-\frac{1}{2}\mu^2 A_\mu A^\mu,
\end{eqnarray}
which leads to the Proca equation~\cite{PhysRevD.96.035019}: 
\begin{eqnarray}
g^{ab}\nabla_a \nabla_b A^\nu =\mu^2 A^\nu.
\end{eqnarray}
Similar to the scalar case, when the Compton wavelength of the vector field exceeds the size of the black hole, the system exhibits non-relativistic, hydrogen-like bound states. The vector field can be written as
\begin{eqnarray}
    A^\nu(t,r,\theta,\phi)=\frac{1}{\sqrt{2\omega}}\big[\Psi^\nu(r,\theta,\phi)e^{-i\omega t}+\text{c.c.} \big].
\end{eqnarray}
After imposing the Lorentz gauge condition $\nabla_\mu A^\mu=0$, the time component $A^0$ is determined and the spatial components of the wave function can be expressed as
\begin{eqnarray}
    \Psi_i&=&R^{nl}(r)Y^{l,jm}_i(\theta,\phi)\nonumber\\
&=&R^{nl}(r)\sum^l_{m_l=-l}\sum^1_{m_s=-1}\braket{(1, m_s),(l, m_l)\mid j, m}\nonumber\\
&&\times\xi^{m_s}_i Y^{lm_l}(\theta,\phi),
\end{eqnarray}
where $Y^{l,jm}_i(\theta, \phi)$ are vector spherical harmonics that encode the coupling between orbital angular momentum and spin and are constructed using Clebsch-Gordan coefficients. The spin basis vectors are defined as
\begin{eqnarray}
    \xi^0=\hat{z},\ \xi^{\pm 1}=\mp \frac{1}{\sqrt{2}}(\hat{x}\pm i \hat{y}). 
\end{eqnarray}
Here, $n$ is the principal quantum number, $l$ is the orbital angular momentum ($l=0,1,2,\dots$), $j$ is the total angular momentum ($j=l-1, l, l+1$; with $j=0,1$ for $l=0$), and $m$ is the magnetic (azimuthal) quantum number ranging from $-j$ to $j$.

\section{Perturbation from companion}

Gravitational perturbations induce overlaps between bound states $\ket{\psi_{a}}\equiv\ket{\psi_{n_a l_a m_a}}$ (or $\ket{\psi_{n_a l_a j_a m_a}}$ for the vector case) and $\ket{\psi_{b}}\equiv\ket{\psi_{n_b l_b m_b}}$ (or $\ket{\psi_{n_b l_b j_b m_b}}$ for the vector case).
The level mixing for scalar boson has been studied in~\cite{Baumann_2019,PhysRevD.101.083019}. In the present work, we focus on evaluating the corresponding effects for the vector boson cloud. The gravitational perturbation induces overlaps between different bound states of the form
\begin{eqnarray}\label{eq:mixing element}
\bra{\psi_{a}}\delta V_*\ket{\psi_{b}}&=
    -&\sum_{l_*\geqslant 2}\sum_{\left|m_*\right|\leqslant l_*}M_* \mu\frac{4\pi}{2l_*+1}\nonumber\\
    &&\times Y^{*}_{l_* m_*}(\Theta_*,\Phi_*)\times I_r \times I_\Omega,
\end{eqnarray}
with $I_r$ and $I_\Omega$ are integrals
\begin{eqnarray}\label{eq:Ir}
    I_r&\equiv &\int_0^{R_*} dr \Big(\frac{r^{2+l_*}}{R^{l_*+1}_*} R_{n_a l_a}R_{n_b l_b}\Big)\nonumber\\
    &&+\int_{R_*}^{\infty} dr \Big(\frac{R_*^{l_*}}{r^{l_*-1}} R_{n_a l_a}R_{n_b l_b}\Big)\nonumber,\\
    I_\Omega&\equiv& \int d\Omega  Y_{l_* m_*}(\theta,\phi)\mathbf{Y}^*_{l_a,j_a m_a}\cdot \mathbf{Y}_{l_b, j_b m_b}.
\end{eqnarray}
To obtain an effective integral, the associated states of the primary are restricted~\cite{PhysRevD.101.083019},
\begin{eqnarray}
    &(1)&\ \mid l_a-l_b \mid \leq l_* \leq l_a+l_b,\nonumber\\
    &(2)&\ \mid j_a-j_b \mid \leq l_* \leq j_a+j_b,\nonumber\\
    &(3)&\ m_* + m_a-m_b=0,\nonumber\\
    &(4)&\ l_a+l_b+l_*=2p ,\ \text{for}\ p\in\mathbb{Z}.
\end{eqnarray}

In this context, we consider a primary black hole surrounded by a vector cloud occupying its $\ket{\psi_{1011}}$ state. According to these selection rules, the closest states that can mix with it are $\ket{\psi_{321-1}}$ and $\ket{\psi_{3210}}$.
Using the basis representation $\ket{1011} = (1, 0, 0)^T$,  $\ket{3210} = (0, 1, 0)^T$, and $\ket{321-1} = (0, 0, 1)^T$, the Hamiltonian matrix elements for the vector field, under gravitational perturbation induced by the companion, can be written as
\begin{eqnarray}
H=H_0+H_1+H_2.
\end{eqnarray}
Here, $H_0$ represents the eigenfrequencies of the unperturbed vector states at different energy levels, while $H_1$ and $H_2$ arise from gravitational perturbations. The matrix $H_1$ contains only diagonal elements, corresponding to the energy shifts of the individual levels, given by $\bra{a}V_*\ket{a}$. In contrast, the off-diagonal elements in $H_2$, given by $\bra{a}V_*\ket{b}$, induce mixing between states of different energy levels .
\begin{widetext}
    \begin{eqnarray}
H_0=
\left(\begin{matrix}
   \big(1-\frac{\alpha^2}{2}-\frac{35 \alpha^4}{24}+\frac{8 \tilde{a} \alpha^5}{3}\big) \mu & 0 & 0 \\
   0 & \big(1-\frac{\alpha^2}{18}-\frac{109\alpha^4}{3240}\big) \mu & 0\\
   0 & 0 & \big(1-\frac{\alpha^2}{18}-\frac{109\alpha^4}{3240}-\frac{4 \tilde{a} \alpha^5}{405} \big) \mu
\end{matrix}\right),
\end{eqnarray}
\begin{eqnarray}
H_1=
\left(\begin{matrix}
   0 & 0 & 0 \\
   0 & -\frac{1}{10}\eta_0 f_0(\Theta_*) & 0\\
   0 & 0 & \frac{1}{20}\eta_0 f_0(\Theta_*)
\end{matrix}\right),
\end{eqnarray}
\begin{eqnarray}
H_2=
\left(\begin{matrix}
   0& -\frac{1}{10}\eta_1 f_1(\Theta_*)e^{\mp i\Omega t}  & -\frac{1}{10\sqrt{2}}\eta_1 f_2(\Theta_*)e^{\mp 2i\Omega t} \\
   -\frac{1}{10}\eta_1 f_1(\Theta_*)e^{\pm i\Omega t} & 0 & -\frac{1}{10\sqrt{2}}\eta_0 f_1(\Theta_*)e^{\mp i \Omega t}\\
  -\frac{1}{10\sqrt{2}}\eta_1 f_2(\Theta_*)e^{\pm 2i\Omega t} & -\frac{1}{10\sqrt{2}}\eta_0 f_1(\Theta_*)e^{\pm i \Omega t} & 0
\end{matrix}\right).
\end{eqnarray}
\end{widetext}
Here, $\eta_0$ and $\eta_1$ characterize the strength of the effect as a function of the binary separation. At leading order, corresponding to the quadrupole term $l_*=2$, the diagonal element is given by
\begin{eqnarray}
\eta_0 
  &=& q M \mu \times I_r \nonumber\\
  &=& \frac{q\, e^{-\tfrac{2R_*\alpha^2}{3M}}}{2187\, M^5 R_*^3 \alpha^3}
     \Big[
       275562 M^7 \!\Big(e^{\tfrac{2R_*\alpha^2}{3M}} - 1 \Big)
       \nonumber\\
  &\quad& -183708 M^6 R_*\alpha^2
       -61236 M^5 R_*^2 \alpha^4 \nonumber\\
  &\quad&
       -13608 M^4 R_*^3 \alpha^6
       -2268 M^3 R_*^4 \alpha^8
       -297 M^2 R_*^5 \alpha^{10}\nonumber\\
  &\quad&       -30 M R_*^6 \alpha^{12}
       -2 R_*^7 \alpha^{14}
     \Big],
\end{eqnarray}
with $f_0$ as a function of the inspiral angle 
\begin{eqnarray}
f_0(\Theta_*)=3\cos^2(\Theta_*)-1.
\end{eqnarray}
While for off-diagonal terms, the strength function representing the effect of separation is
\begin{eqnarray}
\eta_1&=&qM\mu\times I_r\nonumber\\
&=&\sqrt{\frac{5}{6}}e^{-\frac{4R_*\alpha^2}{3M}}q\Big[2187(-1+e^{\frac{4R_*\alpha^2}{3M}})M^5\nonumber\\
&&-2916M^4 R_*\alpha^2-1944M^3 R_*^2\alpha^4\nonumber\\
&&-864M^2 R_*^3 \alpha^6-288M R_*^4 \alpha^8\nonumber\\
&&-64 R_*^5\alpha^{10}\Big]/1152M^3 R_*^3\alpha^3,
\end{eqnarray}
with
\begin{eqnarray}
&f_1(\Theta_*)=3\cos(\Theta_*)\sin(\Theta_*),&\nonumber\\
&f_2(\Theta_*)=3\sin^2(\Theta_*).&
\end{eqnarray}

We assume that the companion black hole lies in the equatorial plane of the primary black hole, $\Theta_*=\pi/2$. In this configuration, the angular functions $f_{|m_*|}(\Theta_*)$ simplify to $f_0=-1, f_1=0, f_2=3$. Consequently, the perturbation energy vanishes for the state $\ket{3210}$, and we focus on the energy level mixing between $\ket{1011}$ and $\ket{321-1}$.
Working in the interaction picture
\begin{eqnarray}
i\frac{d}{dt}\ket{\psi}_I=H_I \ket{\psi}_I,
\end{eqnarray}
labeling $\ket{\psi_{g}}\equiv \ket{1011}$ and $\ket{\psi_{d}}\equiv \ket{321-1}$, 
the state in the interaction picture is described as
\begin{eqnarray}
\ket{\psi_I(t)}=c_g(t)\ket{\psi_{g}}+c_{d}(t)\ket{\psi_{d}},
\end{eqnarray}
where $g$ and $d$ represent ``growing" and ``decaying", respectively. The schr\"{o}dinger equation implies 
\begin{eqnarray}
&&i\frac{d}{dt}
\begin{pmatrix}
c_g(t)\\
c_{d}(t)
\end{pmatrix}
=\\&&
\begin{pmatrix}
  0 &  -\frac{3\eta_1 e^{-2i(\pm\Omega-\epsilon)t}}{10\sqrt{2}}\\
  -\frac{3\eta_1 e^{+2i(\pm\Omega-\epsilon)t}}{10\sqrt{2}}  & 0 
\end{pmatrix}
\begin{pmatrix}
   c_g(t)\\
   c_{d}(t)
\end{pmatrix},\nonumber
\end{eqnarray}
where the energy gap
\begin{eqnarray}\label{resonance point}
\epsilon=-\frac{2}{9}\alpha^2 \mu.
\end{eqnarray}
Starting with all vectors occupying the growing mode, we set $c_g(0)=1$ and $c_d(0)=0$, subject to the normalization condition
\begin{eqnarray}
|c_g(t)|^2+|c_d(t)|^2=1,
\end{eqnarray}
we find 
\begin{eqnarray}
c_g(t)&=&\frac{e^{i(\epsilon \mp \Omega)t}}{2\Delta_R}[(\Delta_R+\epsilon \mp \Omega)e^{-i\int\Delta_R dt'}\nonumber\\
&&+(\Delta_R-\epsilon \pm \Omega)e^{+i\int\Delta_R dt'}],\\
c_d(t)&=&\frac{-\frac{3}{10\sqrt{2}}\eta_1 e^{-i(\epsilon \mp \Omega)t}}{2\Delta_R}[e^{-i\int\Delta_R dt'}-e^{+i\int\Delta_R dt'}],\nonumber
\end{eqnarray}
with
\begin{eqnarray}
\Delta_R \equiv \sqrt{\Big(-\frac{3}{10\sqrt{2}}\eta_1\Big)^2+(\epsilon \mp \Omega)^2} \,.
\end{eqnarray}
The occupation probability of the decaying mode is proportional to 
\begin{eqnarray}
|c_d(t)|^2&=&\frac{-\frac{3}{10\sqrt{2}}\eta_1 e^{-i(\epsilon \mp \Omega)t}}{2\Delta_R}\big( e^{-i\int\Delta_R dt'}-e^{+i\int\Delta_R dt'} \big)\nonumber\\
&&\times \frac{-\frac{3}{10\sqrt{2}}\eta_1 e^{i(\epsilon \mp \Omega)t}}{2\Delta_R}\big( e^{i\int\Delta_R dt'}-e^{-i\int\Delta_R dt'} \big) \nonumber\\
&=&\bigg[ 1-(\frac{\epsilon\mp \Omega}{\Delta_R})^2 \bigg] \sin^2 \bigg[\int_{t_0}^{t}dt^\prime \Delta_R(t^\prime) \bigg].
\end{eqnarray}
Note that $\Delta_R$ varies during the inspiral, and therefore the phase of the oscillations is expressed as an integral over time. 
Since the energy level $\ket{321-1}$ satisfies $\omega_{321-1}>\omega_{1011}$, we have $\epsilon<0$, and the resonance occurs only for counter-rotating orbits.
The corresponding binary separation is given by
\begin{eqnarray}
R_*=2.72\alpha^{-2}(1+q)^{1/3}r_g=2.72(1+q)^{1/3}r_b .
\end{eqnarray}

\section{Extension to Molecular orbitals with $l>1$}\label{more n}

In the previous analysis, we restricted our consideration to molecular orbitals with $l\leq 1$. For $\sigma$ orbitals, all participating atomic orbitals necessarily satisfy $l\leq 1$. However, for $\pi$ orbitals the companion black hole's $3d_{xy}$ state also meets the required symmetry conditions. Nevertheless, the occupation probability of the associated atomic orbital $\psi_{32-2}$ is found to be much lower than that of the $l\leq 1$ states, and its decay rate is lower by several orders of magnitude. Consequently, the contribution from such higher-$l$ orbitals can be neglected in the analysis of boson cloud depletion. This also applies to other molecular orbitals with $l\leq 1$.

We calculate a specific example for $q=1.5$ and $\alpha=0.1$.  Among the companion black hole's atomic states, the one with $n=3$ has an energy most closely matching that of the primary black hole for this mass ratio. Therefore, we consider the companion black hole's atomic states with principal quantum numbers ranging from $n=1$ to $n=3$. 
\begin{eqnarray}
    \ket{\sigma_s^a}&=& f_{a0}\ket{\psi^1_{2p_x}}+f_{a1}\ket{\psi^1_{1s}}+f_{a2}\ket{\psi^2_{2s}}\nonumber\\
    &&+f_{a3}\ket{\psi^2_{3s}}+f_{a4}\ket{\psi^2_{2p_x}}+f_{a5}\ket{\psi^2_{3p_x}},
\end{eqnarray}
while for the $\ket{\pi_s}$ orbitals an additional $d$ state needs to be accounted for,
\begin{eqnarray}
    \ket{\pi_s^b}&=& c_{b1}\ket{\psi^1_{2p_y}}+c_{b2}\ket{\psi^2_{2p_y}}\nonumber\\
    &&+c_{b3}\ket{\psi^2_{3p_y}}+c_{b4}\ket{\psi^2_{3d_{xy}}}. 
\end{eqnarray}

\begin{figure}[htbp]
    \centering
    \includegraphics[width=1\columnwidth]{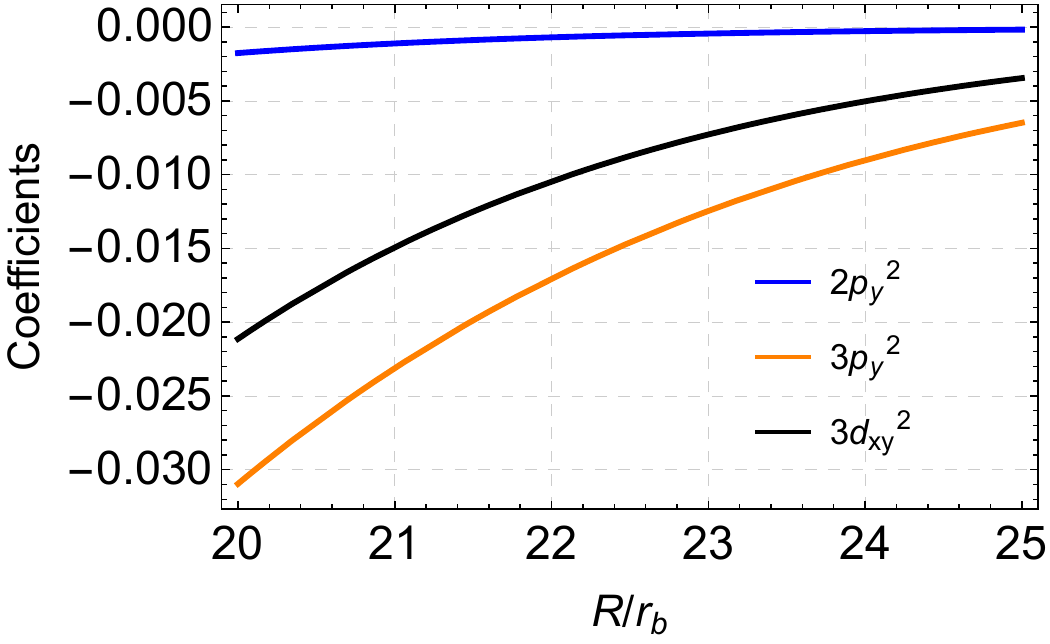}
    \caption{Coefficients of the dominant $\pi_s$ molecular orbits. For mass ratio $q=1.5$ and $\alpha=0.1$.}
    \label{fig:scalar_y_coef}
\end{figure}

\begin{figure}[htbp]
    \centering
    \includegraphics[width=1\columnwidth]{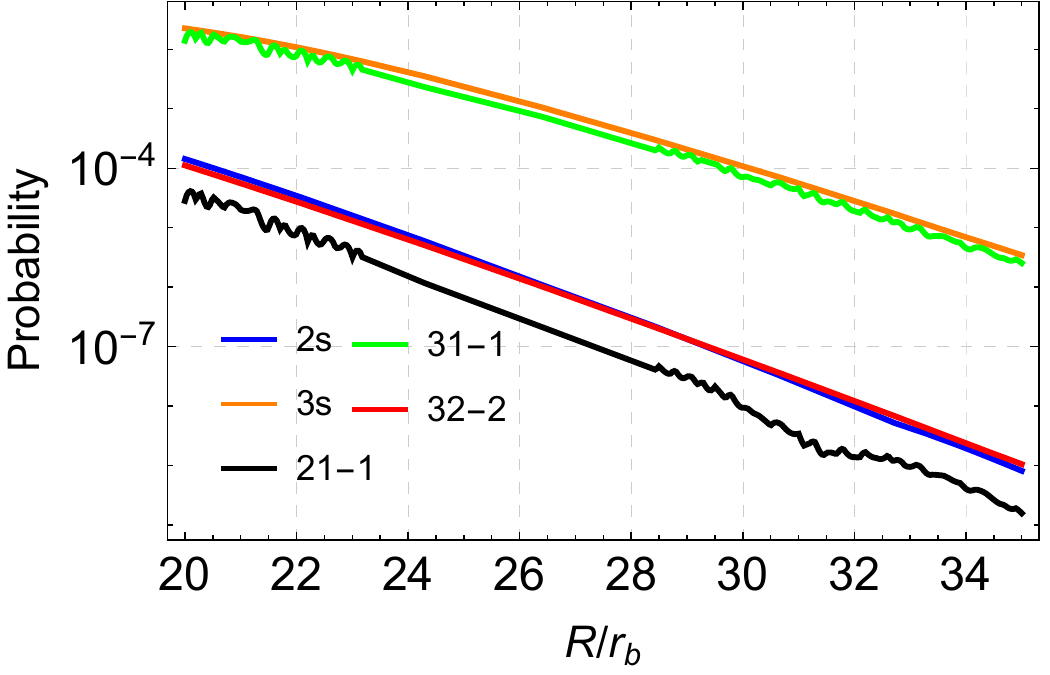}
    \caption{Occupation probabilities of the decaying modes of the companion black hole. For mass ratio $q=1.5$ and $\alpha=0.1$.}
    \label{fig:scalar more n occupation density}
\end{figure}

In Fig.~\ref{fig:scalar_y_coef}, we show the atomic coefficients of one of the $\pi$ orbitals. These coefficients characterize the strength of the bonding between the two black holes. For a mass ratio of $q=1.5$, the states with a principal quantum number of $n=3$ have energies closer to that of the primary state. Consequently, the coefficients with the $n=3$ states are larger than those of the $n=2$ states, indicating a stronger bonding effect. Furthermore, we can see that the absolute value of the atomic coefficient corresponding to the $3p_y$ orbital is larger than that of the $3d_{xy}$ orbital. In addition, since the $\sigma$ orbital involves the $\psi_{31-1}$ atomic state but not the $\psi_{32-2}$ state, the occupation probability of the $\psi_{31-1}$ state is significantly higher than that of $\psi_{32-2}$, as shown in Fig.~\ref{fig:scalar more n occupation density}.

\begin{figure}[htbp]
    \centering
    \includegraphics[width=1\columnwidth]{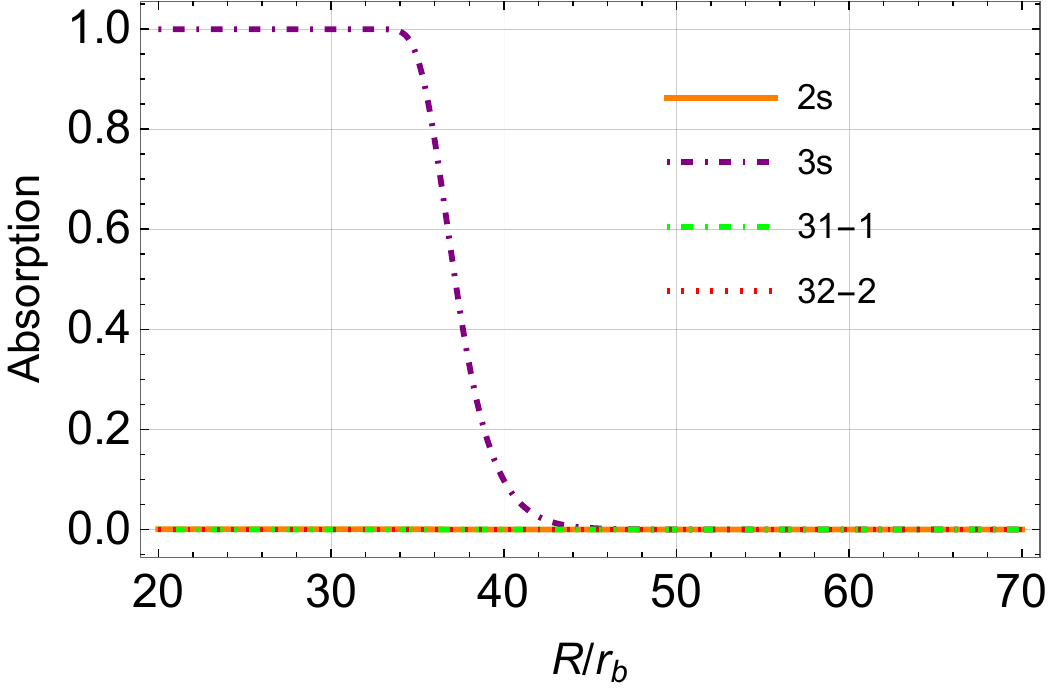}
    \caption{Absorption percentage of each decaying modes of the companion. For mass ratio $q=1.5$ and $\alpha=0.1$.}
    \label{fig:scalar_more_depletion}
\end{figure}

Regarding the occupation probabilities, we evaluate the fraction of the cloud depleted through these decaying modes. In Fig.~\ref{fig:scalar_more_depletion}, we present the absorption of the boson cloud into the corresponding atomic states. Although the states with angular momentum quantum number $l=1$ exhibit higher occupation probabilities compared to the $s$ states, it is the $s$ states that dominate the overall depletion due to their much larger decay rates. In contrast, the $\ket{\psi^2_{32-2}}$ state, which possesses both a low occupation probability and a small depletion rate, makes only a negligible contribution.

\section{Non-adiabatic evolution}\label{Non-adiabatic evolution}

Considering that non-adiabatic transitions between orbitals occur on timescales much longer than those of free evolution terms, we factor out the rapidly oscillating phase and absorb the orbital transitions into the time-dependent coefficients $a_n(t)$. This leads to the following expansion of the quantum state:
\begin{eqnarray}
    \ket{\psi(t)}=\sum_n a_n(t) e^{-i\int_{-\infty}^t E_n d\tau}\ket{\psi_n(t)}.
\end{eqnarray}
Substituting into the time-dependent shr\"{o}dinger equation,
\begin{eqnarray}
    i\frac{d}{dt}\ket{\psi(t)}=\hat{H}(t)\ket{\psi(t)},
\end{eqnarray}
we obtain
\begin{eqnarray}
    &&\sum_n(\dot{a}_ne^{-i\int_{-\infty}^t E_n d\tau}\ket{\psi_n}+a_n E_n\ket{\psi_n}\nonumber\\
    &&+a_n e^{-i\int_{-\infty}^t E_n d\tau}\ket{\dot{\psi}_n})
    =\sum_n a_n E_n\ket{\psi_n}.
\end{eqnarray}
Multiplying both sides from the left by $\bra{\psi_n}$, we find
\begin{eqnarray}
    &&\dot{a}_ne^{-i\int_{-\infty}^t E_n d\tau}+a_n E_n+a_n e^{-i\int_{-\infty}^t E_n d\tau}\braket{\psi_n\mid\dot{\psi}_n}\nonumber\\
    &&+\sum_{k\neq n}a_n e^{-i\int_{-\infty}^t E_n d\tau}\braket{\psi_n\mid\dot{\psi}_k}=a_n E_n.
\end{eqnarray}
Since $\braket{\psi_n\mid\dot{\psi}_n}$ is purely imaginary, it can be removed through an appropriate phase redefinition. Thus, the only physically relevant contribution comes from the off-diagonal terms,
\begin{eqnarray}
    \frac{\partial a_n}{\partial t}&=&-\sum_{k\neq n}a_k e^{-i\int_{-\infty}^t(E_k-E_n)d\tau}\braket{\psi_n\mid\dot{\psi}_k}\nonumber\\
    &=&-\sum_{k\neq n}\frac{a_k e^{-i\int_{-\infty}^t(E_k-E_n)d\tau}}{E_k-E_n}\bra{\psi_n}\frac{\partial \hat{H}}{\partial t}\ket{\psi_k}.
\end{eqnarray}
To illustrate the non-adiabatic evolution of the boson cloud, we calculate the off-diagonal terms. It can be divided into three parts. The first component arises from the cross term between the $\sigma$-type molecular orbitals:
\begin{eqnarray}\label{eq:SigmaCross}
    &&\bra{\sigma^i}\frac{\partial \hat{H}}{\partial t}\ket{\sigma^j}=f_{i0}f_{j0}\bra{\psi^1}\frac{\partial \hat{H}}{\partial t}\ket{\psi^1}\nonumber\\
    &&+\sum_{n=1}^N (f_{i0}f_{jn}+f_{j0}f_{in})\bra{\psi^1}\frac{\partial \hat{H}}{\partial t}\ket{\psi^2_{ns}}\nonumber\\
    &&+\sum_{n=2}^N (f_{i0}f_{j(N+n-1)}+f_{j0}f_{i(N+n-1)})\bra{\psi^1}\frac{\partial \hat{H}}{\partial t}\ket{\psi^2_{np_x}}\nonumber\\
    &&+\sum_{n=1}^N \sum_{m=2}^N(f_{in}f_{j(N+m-1)}+f_{jn}f_{i(N+m-1)})\bra{\psi^2_{ns}}\frac{\partial \hat{H}}{\partial t}\ket{\psi^2_{mp_x}}\nonumber\\
    &&+\sum_{n=1}^N \sum_{m=1}^N f_{in}f_{jm}\bra{\psi^2_{ns}}\frac{\partial \hat{H}}{\partial t}\ket{\psi^2_{ns}}\nonumber\\
    &&+\sum_{n=2}^N \sum_{m=2}^N f_{i(N+n-1)}f_{j(N+m-1)}\bra{\psi^2_{np_x}}\frac{\partial \hat{H}}{\partial t}\ket{\psi^2_{mp_x}}
\end{eqnarray}
For the scalar boson case, $\psi^1$ in Eq.~\eqref{eq:SigmaCross} represents the atomic orbital $\psi^1_{2p_x}$ of the primary black hole; whereas for the vector boson, it corresponds to $\psi^1_{1011}$. It is also convenient to express the $\pi_s$ orbitals in the same form as above. 
\begin{eqnarray}
    &&\bra{\pi^\mu}\frac{\partial \hat{H}}{\partial t}\ket{\pi^\nu}=g_{\mu0}g_{\nu0}\bra{\psi^1_{2p_y}}\frac{\partial \hat{H}}{\partial t}\ket{\psi^1_{2p_y}}\nonumber\\
    &&+\sum_{n=2}^N (g_{\mu0}g_{\nu n}+g_{\nu 0}g_{\mu n})\bra{\psi^1_{2p_y}}\frac{\partial \hat{H}}{\partial t}\ket{\psi^2_{np_y}}\nonumber\\
    &&+\sum_{n=2}^N \sum_{m=2}^N g_{\mu n}g_{\nu m}\bra{\psi^2_{np_y}}\frac{\partial \hat{H}}{\partial t}\ket{\psi^2_{mp_y}}. 
\end{eqnarray}
It can also be shown that the $\sigma_s$ orbitals do not couple to the $\pi_s$ orbitals, namely,
\begin{eqnarray}
    \bra{\sigma^i} \frac{\partial \hat{H}}{\partial t}\ket{\pi^\mu}=0.
\end{eqnarray}

For equal-mass binaries, the scalar cloud evolves adiabatically, whereas in unequal-mass systems, bosons can transition between different orbitals through non-adiabatic processes during the evolution. We examine how these non-adiabatic transitions affect the occupation probabilities. Our calculations show that the transitions resulting from non-adiabatic evolution are extremely weak, indicating that the scalar cloud remains largely in its initial molecular orbital.

As a representative example, we take $q=0.46$ and $\alpha=0.1$. In this case, the molecular orbital with the highest occupation becomes nearly degenerate in energy with another orbital at a certain separation, which enhances the non-adiabatic transitions. However, the resulting effect is still found to be very weak. 
\begin{figure}[htbp]
    \centering
    \includegraphics[width=1\columnwidth]{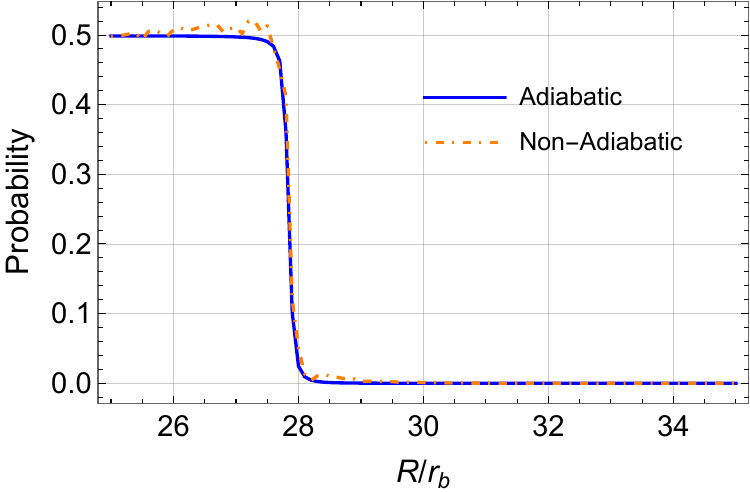}
    \caption{Occupation probabilities of the scalar cloud in the companion state $\ket{\psi_{1s}^2}$ for $q=0.46$, $\alpha=0.1$. The blue solid line shows the adiabatic evolution, whereas the orange dashed line includes the effect of non-adiabatic transitions. }
    \label{fig:scalar non-adia compare}
\end{figure}

\section{Special mass ratio}\label{appendix special mass ratio}

For both scalar and vector clouds, there exist specific mass ratios of the binary black hole system at which cloud depletion occurs at larger separations compared to nearby mass ratios.
It is evident from Fig.~\ref{fig:scalar total depletion} that there exist certain mass ratios that show significantly larger depletion distances compared to nearby mass ratios. This occurs when the binding energy of $\psi^1_{2p_x}$ matches that of the atomic state of the companion. The mass ratios corresponding to cloud depletion at larger separations are found to be approximately $q=0.5$, $q=1$, and $q=1.5$. At these mass ratios, the energy of the primary black hole's $\psi^1_{2p_x}$ atomic state approaches that of the $n=1, 2, 3$ atomic states of the companion black hole, respectively.

\begin{figure}[htbp]
    \centering
   \includegraphics[width=1\columnwidth]{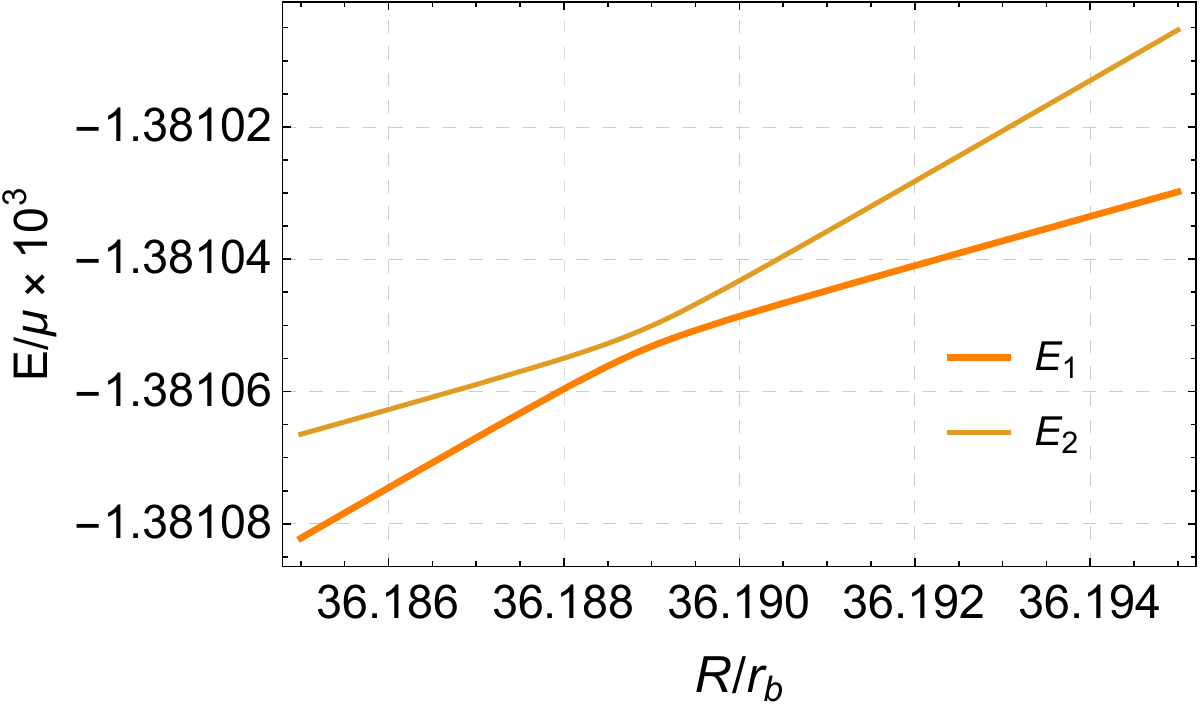}
    \caption{Energy eigenvalues of the near-degenerate states. We choose $q = 0.47$ and $\alpha=0.1$}
    \label{fig:scalar_energy_047}
\end{figure}

Next, we consider a representative case with mass ratio $q=0.47$, in which the energy of the companion atomic state $\ket{\psi^2_{1s}}$ is closest to that of the primary black hole. 
We present the energies of two molecular orbitals that correspond to the atomic orbitals $\psi^1_{2p_x}$ and $\psi^2_{1s}$ at infinity, respectively. At an orbital separation of about $36r_b$, as shown in Fig.~\ref{fig:scalar_energy_047}, the two molecular orbitals are nearly degenerate in energy, which significantly enhances the non-adiabatic coupling between them.

\begin{figure}[htbp]
    \centering
    \includegraphics[width=1\columnwidth]{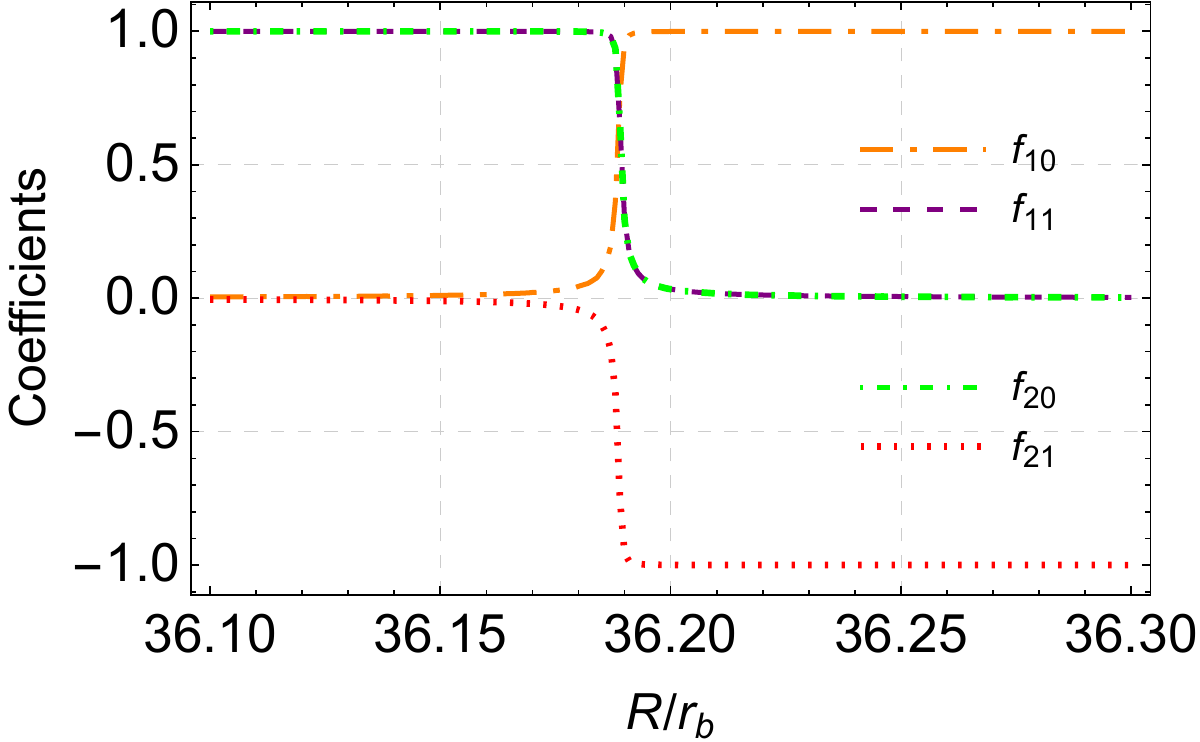}
    \caption{Coefficients of the associated atomic states. We choose $q=0.47$ and $\alpha=0.1$}
    \label{fig:scalar_c1coef_047}
\end{figure}

Figure~\ref{fig:scalar_c1coef_047} illustrates the coefficients of $\psi^1_{2p_x}$ and $\psi^2_{1s}$ of these two molecular orbitals. We can see that the coefficients of the two molecular orbitals change significantly. For the molecular orbital that corresponds to $\psi^1_{2p_x}$ at infinity, the coefficient $f_{10}$ gradually decreases to nearly zero, while $f_{11}$ increases and approaches unity. In contrast, the coefficients of the molecular orbitals corresponding to $\psi^2_{1s}$ exhibit the opposite behavior.

When the binary black holes are widely separated at $R_*=80 r_b$, the bosons predominantly occupy the isolated orbit $\ket{\psi^1_{211}}$, which corresponds to the first molecular orbital. As the separation decreases toward the near-degeneracy point, the narrow energy gap facilitates non-adiabatic transitions between molecular orbitals.  However, in this regime, most of the cloud still remains in the initial molecular orbital, with only a small fraction transitioning to other molecular orbitals.
After passing through the near-degenerate point, the small deviation of $F_a$ and larger variation in $f_{11}$ indicate that the occupation probability of $\psi^2_{1s}$ increases dramatically. The bosons occupying $\psi^2_{1s}$ are then absorbed by the companion black hole, leading to a rapid transfer and depletion of the boson cloud.

\section{Arbitrary spin orientation}
\label{Appendix:spin_orientation}

When the spin orientation of the companion is altered, the wave functions are changed accordingly. This transformation can be represented using the Wigner $D$-matrix, which performs the full Euler rotation sequence $(\zeta,\beta,\gamma)$ on the spherical harmonic basis. The rotation about the $z$-axis affects the $p_x$ and $p_y$ states only through a phase factor, without changing their physical overlap. Consequently, the rotation angle $\gamma$ has no physical significance in this context. Therefore, it is appropriate to disregard $\gamma$ when analyzing the overlap and focus instead on the nontrivial effects of the $\zeta$ and $\beta$ rotations.
Rotations about the $x$- and $y$-axes are generated by the orbital angular momentum operators $J_x$ and $J_y$. 
\begin{eqnarray}
    \hat{R}^1_x(\zeta)&=&e^{-i\zeta J_x/\hbar }\nonumber\\
    &=&\exp\left(-i\zeta\left(\begin{matrix}
   0 & 1/\sqrt{2} & 0 \\
   1/\sqrt{2}& 0 & 1/\sqrt{2}\\
   0 & 1/\sqrt{2} & 0
\end{matrix}\right)\right),\nonumber\\
    \hat{R}^1_y(\beta)&=&e^{-i\beta J_y/\hbar }\nonumber\\
    &=&\exp\left(-i\beta\left(\begin{matrix}
   0 & -i/\sqrt{2} & 0 \\
   i/\sqrt{2}& 0 & -i/\sqrt{2}\\
   0 & i/\sqrt{2} & 0
\end{matrix}\right)\right).
\end{eqnarray}

Diagonalizing the above matrix allows us to extract the physically meaningful transformation components. The Wigner-$D$ matrix corresponding to a rotation about the $y$-axis by an angle $\beta$ in the $l=1$ subspace can be written as
\begin{eqnarray}
    D^1_y(\beta)=U e^{-i\beta\Lambda_y/\hbar}U^\dagger,
\end{eqnarray}
where $\Lambda_y=\text{diag}(\hbar,0,-\hbar)$ and $U$ is the unitary matrix
\begin{eqnarray}
    U=\left(\begin{matrix}
   1/2 & 1/\sqrt{2} & 1/2 \\
   i/\sqrt{2}& 0 & -i/\sqrt{2}\\
   1/2 & 1/\sqrt{2} & 1/2
\end{matrix}\right).
\end{eqnarray}
The corresponding Wigner-$D$ matrix for a rotation about the $y$-axis by an angle $\beta$ then takes the explicit form:
\begin{eqnarray}
    D^1_y(\beta)=\frac{1}{2} \begin{pmatrix}
   1+\cos\beta & -\sqrt{2} \sin \beta & 1-\cos\beta \\
   \sqrt{2} \sin \beta & 2 \cos \beta & -\sqrt{2} \sin\beta \\
   1-\cos\beta & \sqrt{2} \sin \beta & 1+\cos\beta
\end{pmatrix}.
\end{eqnarray}
Similarly, the rotation matrix corresponding to a rotation about the $x$-axis by an angle $\zeta$ is
\begin{eqnarray}
\label{eq: Dx}
D^1_x(\zeta)=\frac{1}{2}\left(\begin{matrix}
   1+\cos\zeta & -i\sqrt{2}\sin\zeta & \cos\zeta-1 \\
   -i\sqrt{2}\sin\zeta& 2\cos\zeta & -i\sqrt{2}\sin\zeta\\
   \cos\zeta-1 & -i\sqrt{2}\sin\zeta & 1+\cos\zeta
\end{matrix}\right).
\end{eqnarray}

The recombination of spherical harmonics indicates that the state $\psi_{np_x}(\mathbf{r})$ can be represented as a linear combination of the real co-rotating wave functions. 
The final form of the rotated wave function is
\begin{eqnarray}
    \psi_{np_x}(\mathbf{r})=&\cos\beta \, \psi_{np_x}(\mathbf{r'})+\sin\beta\sin\zeta \, \psi_{np_y}(\mathbf{r'})\nonumber\\
    &-\cos\zeta\sin\beta \, \psi_{np_z}(\mathbf{r'}).
\end{eqnarray}
Due to the rotation of the companion black hole's spin direction, the newly formed atomic orbital $\psi^2_{np_x}(\mathbf{r})$ differs from that in the spin-parallel configuration. The key question is whether this newly combined atomic orbital changes its wavefunction overlap with the atomic orbital of the primary black hole, thereby affecting the formation of the molecular orbitals. 
To compute the molecular orbitals formed by the primary and companion black holes, we need the overlap between their respective atomic orbitals, for instance, $\braket{\psi^1_{2p_x}\mid \psi^2_{np_x}(\mathbf{r})}$.  The overlap of the wave functions can be expressed as
\begin{eqnarray}
    &&\braket{\psi^1_{2p_x}\mid\psi^2_{n p _x}(\mathbf{r})}=\cos\beta \braket{\psi^1_{2p_x}\mid\psi^2_{np_x}(\mathbf{r'})}+\sin\beta\sin\zeta\nonumber\\
    &&\times\braket{\psi^1_{2p_x}\mid\psi^2_{np_y}(\mathbf{r'})}-\cos\zeta\sin\beta\braket{\psi^1_{2p_x}\mid\psi^2_{np_z}(\mathbf{r'})}
\end{eqnarray}
Since $\psi^2_{np_x}(\mathbf{r})$ is expressed as a linear combination of the companion’s atomic orbitals, this calculation effectively reduces to evaluating the set of overlaps $\braket{\psi^1_{2p_x} \mid \psi^2_{np_x}(\mathbf{r'})}$, $\braket{\psi^1_{2p_x} \mid \psi^2_{np_y}(\mathbf{r'})}$ and $\braket{\psi^1_{2p_x} \mid \psi^2_{np_z}(\mathbf{r'})}$. Therefore, it is necessary to determine explicitly how the wave functions $\psi_{np_x}(\mathbf{r'})$, $\psi_{np_y}(\mathbf{r'})$, and $\psi_{np_z}(\mathbf{r'})$ transform under the Euler rotations. 
This transformation is represented by a three-dimensional rotation matrix acting on the spatial coordinates of the wave functions.
Thus, the angular coordinates $\theta$ and $\phi$ in the rotated frame $\mathbf{r'}$ are related to the initial coordinates $\mathbf{r}$ through the following transformation:
\begin{eqnarray}
    \left(\begin{matrix}
   x'=r'\sin\theta'\cos\phi'\\
   y'=r'\sin\theta'\sin\phi'\\
   z'=r'\cos\theta'
\end{matrix}\right)
=
R_{x}(\zeta)R_{y}(\beta)\left(\begin{matrix}
   r\sin\theta\cos\phi\\
   r\sin\theta\sin\phi\\
   r\cos\theta
\end{matrix}\right),\nonumber\\
\end{eqnarray}
where $R_x(\zeta)$ and $R_y(\beta)$ are rotational matrices.
\begin{eqnarray}
    R_x(\zeta)=\left(\begin{matrix}
   1 & 0 & 0 \\
   0 & \cos\zeta & -\sin\zeta\\
   0 & \sin\zeta & \cos\zeta
\end{matrix}\right),
\end{eqnarray}
and
\begin{eqnarray}
    R_y(\beta)=\left(\begin{matrix}
   \cos\beta & 0 & \sin\beta \\
   0 & 1 & 0\\
   -\sin\beta & 0 & \cos\beta
\end{matrix}\right).
\end{eqnarray}
We now show that the magnitude of the wave function overlap is independent of the rotation angle. Let $\ket{\psi^{2}_{np_x}(\zeta,\beta)}$ denote the state obtained after rotation. One can show that its overlap with the primary state is identical to that of the unrotated state $\ket{\psi^{2}_{np_x}(\mathbf{r})}$.
\begin{widetext}
    \begin{eqnarray}
    \braket{\psi^1_{2p_x}\mid\psi^2_{n p _x}(\zeta,\beta)}&=&\cos\beta \braket{\psi^1_{2p_x}\mid\psi^2_{np_x}(\mathbf{r'})}+\sin\beta\sin\zeta\times\braket{\psi^1_{2p_x}\mid\psi^2_{np_y}(\mathbf{r'})}-\cos\zeta\sin\beta\times\braket{\psi^1_{2p_x}\mid\psi^2_{np_z}(\mathbf{r'})}\nonumber\\
    &=&\cos \beta [\cos \beta\braket{\psi^1_{2p_x}\mid \psi^2_{np_x}(\mathbf{r})}]+\sin\beta \sin\zeta\cdot[\sin\zeta\sin\beta\braket{\psi^1_{2p_x}\mid\psi^2_{np_x}(\mathbf{r})}]\nonumber\\
    &&-\cos\zeta\sin\beta\cdot[-\cos\zeta\sin\beta\braket{\psi^1_{2p_x}\mid\psi^2_{np_x}(\mathbf{r})}]\nonumber\\
    &=&(\cos^2\beta+\cos^2\zeta\sin^2\beta+\sin^2\zeta\sin^2\beta)\braket{\psi^1_{2p_x}\mid\psi^2_{np_x}(\mathbf{r})}\nonumber\\
    &=&\braket{\psi^1_{2p_x}\mid\psi^2_{np_x}(\mathbf{r})}. 
    \end{eqnarray}
\end{widetext}

\bibliography{ref_BosonCloud}

\end{document}